\documentclass[twocolumn]{aastex631}

\accepted{March 16, 2024}

\shorttitle{MIR excess}

\graphicspath{{./}{figures/}}

\begin{document}

\title{Frequencies of warm debris disks based on point source catalogs of {\it Spitzer}, WISE, and {\it Gaia}}

\correspondingauthor{Toshiyuki Mizuki} \email{toshiyuki.mizuki.astr$\_\alpha\tau\_$gmail.com}
\author{Toshiyuki Mizuki}
\affiliation{College of Science, Ibaraki University, Bunkyo 2-1-1, Mito, Ibaraki 310-8512, Japan}
\author{Munetake Momose}
\affiliation{College of Science, Ibaraki University, Bunkyo 2-1-1, Mito, Ibaraki 310-8512, Japan}
\author{Masataka Aizawa}
\affiliation{Tsung-Dao Lee Institute, Shanghai Jiao Tong University, Shengrong Road 520, 201210 Shanghai, China}
\affiliation{Cluster for Pioneering Research, RIKEN, 2-1 Hirosawa, Wako, Saitama 351-0198, Japan}
\author{Hiroshi Kobayashi}
\affiliation{Department of Physics, Nagoya University, Furo-cho, Chikusa-ku, Nagoya, Aichi 464-8602, Japan}

\def\ngaia{25,673,612}
\def\nmir{4,079,230}
\def\ncmd{2,993,308}
\def\nms{2,031,431}
\def\nphot{1,789,617}
\def\nbright{182,586}

\newcommand{\newiset}{15,166}
\newcommand{\newisef}{583}
\newcommand{\neseipm}{575}

\newcommand{\ndwiset}{373}
\newcommand{\ndwisef}{485}
\newcommand{\ndseipm}{255}

\newcommand{\dualexcessa}{35,361}
\newcommand{\dualexcessb}{3,019}
\newcommand{\dualexcessc}{1,055}

\newcommand{\gaiabinary}{279,644}
\newcommand{\nchisq}{2,528,171}

\begin{abstract}
  More than a thousand warm debris disks have been detected as infrared excess at mid-infrared wavelengths, and their frequencies have been obtained for various spectral types of stars.
  However, the dependence of the frequencies on spectral type is still debated because the number of stars with significant and detectable infrared excess is limited.
  Herein, we present the largest systematic search for infrared excess using data from {\it Gaia}, WISE, and {\it Spitzer}.
  We identified $\ndwiset$, $\ndwisef$, and $\ndseipm$-reliable infrared excesses in the mid-infrared archival data at wavelengths of 12, 22, and 24 $\mu$m for WISE/$W3$, $W4$, and {\it Spitzer}/MIPS ch1, respectively.
  Although we confirmed that more massive stars tend to show higher frequencies of debris disks, these disk frequencies are relatively flat for both low- and intermediate-mass stars, with a jump at 7000 K for all three wavelengths.
  Assuming that bright, warm debris disks have lifetimes of a few to several hundred million years, the disk frequency can be understood as the ratio between the timescale and the upper limits of the sample ages.
  We also found that intermediate-mass stars with infrared excess tend to be bluer and fainter along the evolutionary track than those without, implying that massive stars hosting debris disks are relatively young, with an isochronal age of approximately 500 Myr.
  These tendencies are reasonably explained by a standard scenario in which debris disks are likely to be produced by collisions of planetesimals in early stages of stellar evolution, such as the Late Heavy Bombardment. 
\end{abstract}


\section{Introduction}

Debris disks are faint, dusty disks associated with main sequence stars.
Planetary systems form in disks surrounding young stars, called protoplanetary disks. 
In such protoplanetary disks, dust grains evolve into kilometer-sized planetesimals and eventually into planets through aggregation and collisional merging \citep[e.g.,][]{kobayashitanaka21, kobayashitanaka23}. 
Even after the formation of planets, some planetesimals remain in the system as asteroids, similar to the main belt of the Solar System. 
Debris disks comprise such solid materials of various sizes, at least from several $\mu$m-sized dust grains to km-sized planetesimals, serving as indirect signs of planet formation. \par

A fundamental process in the production of dust grains is the collisional cascade. 
In this process, planetesimals are ground into smaller dust grains \citep[e.g.,][]{dohnanyi69, wyatt07, kobayashitanaka10}.
While questions remain open about the evolution of debris disks, i.e., how planetesimals are stirred in a debris disk and how the collisional cascade is ignited \citep[e.g.,][]{wyatt08}, it is thought that the mass of the disks is gradually lost over time as small dust grains resulting from collisional cascades are blown out by stellar radiation pressure or fall to the host stars.
Based on the steady-state collisional cascade model, a swarm of 100 km-sized planetesimals similar to the main belt cannot sustain dust grains detectable within current observational limits for more than 100 Myr \citep[e.g., Equation 3 of][]{ishihara17}.
Only planetary systems that have recently undergone significant collisional fragmentation may have such large populations of dust grains around old host stars \citep[e.g.,][]{genda15, kobayashilohne14}.
Thus, the fraction of stars that are recognized as the site of debris disks is limited. \par

Since the discovery of the debris disk around Vega \citep{aumann84}, observational studies of debris disks have been carried out in two main ways: photometric observations at mid-infrared wavelengths\footnote{The wavelength range designations are set as follows: Optical: 0.4--1 $\mu$m, Near-infrared: 1--3 $\mu$m, and Mid-infrared: 3--40 $\mu$m, although the boundaries depend on the definitions.} and high-contrast imaging at visible and near-infrared wavelengths. 
Dust grains in debris disks, typically sub-$\mu$m in size, have a high scattering efficiency at visible and near-infrared wavelengths, although the blow-out timescales of such small grains by radiation pressure are so short that observable disks are limited.
High-resolution imaging of the scattered light from the central star with large telescopes provide information about the detailed structure of these disks.
Meanwhile, at longer wavelengths, thermal emission from dust disks is observed. 
At mid-infrared wavelengths, the dust scattering efficiency drops sharply while the emission efficiency remains moderate. 
Therefore, at wavelengths corresponding to the dust temperature (as well as to the distance from the central star), the thermal emission from the dust is detected as an excess component on the spectral energy distribution (SED) of the central star.
Using data from the infrared space observatories launched this century ({\it Spitzer,} {\it AKARI}, and WISE: Wide-field Infrared Survey Explorer), the search for excess emission at mid-infrared wavelengths allows statistical studies with a larger sample due to the higher sensitivity for detecting dust disks \citep[e.g.,][]{cotten16}. 
More recently, ALMA imaging observations of debris disks at millimeter and submillimeter wavelengths have also made remarkable progress, revealing dust and gas components at lower temperatures of a cold disk \citep[e.g.,][]{hudges18}
\par 

The evolution of debris disks has been studied with a focus on stars in coeval groups, such as moving groups and open clusters. 
Comparisons of the young groups of different ages suggest that the disk frequencies for early-type and solar-type stars decay significantly on a timescale of approximately 100 Myr to 1 Gyr \citep{siegler07, gaspar09}.
A similar evolution has also been suggested in the detection of disks for low-mass stars. 
\cite{binks17} found 19 infrared excesses out of 100 M dwarfs at WISE/$W4$, but all of the disks are around stars younger than 30 Myr.
Meanwhile debris disk searches without age criteria (i.e., mainly for main sequence stars) find that the frequency of debris disks is speculated to depend on the stellar spectral type.
Although the number of detected disks is very limited, an unbiased all-sky survey using {\it AKARI} has found a possible indication that hotter stars have a higher excess rate \citep{fujiwara13}.
The search for debris disks with WISE has shown that the disk frequencies of the BA-type stars are approximately an order of magnitude higher than those of the FGK-type stars \citep{wu13, patel14}.
However, due to the insufficient number of debris disks, the dependence of the disk frequency on stellar age and mass has not been well understood. 
In particular, disks around low-mass\footnote{Stars with a mass ranging from 0.1 $M_{\odot}$ to 1.5--2 $M_{\odot}$ are categorized as low-mass stars, while those falling below 8 $M_{\odot}$ but exceeding the mass of low-mass stars are referred to as intermediate-mass stars. Within the low-mass star category, a further distinction may be made between solar-type stars and those with even lower masses, typically defined by a boundary of 0.5--0.8 $M_{\odot}$.
} stars have rarely been detected. 
Some young low-mass stars have bright disks, such as AU Mic \citep{kalas04}, but the disk frequency for field low-mass stars could be very low \citep[e.g.,][]{gautier07, avenhaus12}. \par

Debris disks have been detected around various types of host stars.
To avoid the degeneracy between the stellar radius and distance from the Solar System, sources with known parallax have often been employed as targets to search for debris disks in previous studies. 
Before the second {\it Gaia} data release of precise astrometry and photometry for over 1.6 billion stars in the Milky Way \citep{gaiadr2}, parallaxes almost exclusively came from {\it Hipparcos} \citep{hipparcos}.
While many objects, particular low-mass stars, were missed in the {\it Hipparcos} catalog due to limited sensitivity, they have actually been detected by infrared space observatories such as {\it Spitzer} and WISE. 
{\it Hipparcos} parallaxes are limited to sources with $V \sim 10$ mag, corresponding to a 0.5 $M_{\odot}$ main sequence (early M-type) star at $\sim$ 15 pc.
In contrast, infrared observations by WISE/$W3$, $W4$ and {\it Spitzer}/Multiband Imaging Photometer for {\it Spitzer} channel 1 (MIPS ch1 or $M1$) can detect the same object within 150, 30, and 200 parsecs, respectively, according to the detection limits \citep{wise, rieke04}.
{\it Gaia} is capable of measuring parallaxes for sources as faint as $G (\approx V) ~ 20$ mag. \par

This work aims to quantify the frequency of debris disks for main sequence stars as a function of stellar temperature (a proxy for stellar mass on the main sequence), using the data products from the {\it Gaia}, {\it Spitzer,} and WISE surveys.
The data for distant and faint stars in infrared archives are employed, whose detection limits are determined by the sensitivity of the observations for each star rather than by the calibration of the survey \citep[see the concepts of the sensitivity-limited and calibration-limited surveys\footnote{A sensitivity-limited survey is one in which the detectable flux of debris disks is limited by sensitivity, such as exposure time.
On the other hand, a survey whose detectable flux of debris disks is determined by fractional uncertainties, i.e. instrumental and flux calibrations, is called a calibration-limited survey.}, e.g.,][]{wyatt08}.
The remainder of this work is organized as follows:
Section \ref{sec:data} provides details of the point-source catalogs used to characterize the stellar properties and search for infrared excess, and the selection of main sequence star candidates that potentially host debris disks in color-magnitude diagrams.
Section \ref{sec:analysis} describes the use of several catalogs, the characterization of host stars via the fitting of SEDs, and the identification of sources with infrared excess.
Section \ref{sec:correction} presents how our sensitivity-limited survey differs from the calibration-limited surveys of previous studies and the problems in deriving reliable debris disk frequencies.
Section \ref{sec:result} presents the derivation of the debris disk frequencies and comparisons with previous works.
In Section \ref{sec:discussion}, we discuss the dependence on the host stars.
The results are summarized in Section \ref{sec:summary}.

\begin{deluxetable*}{ccc ccc}
  \tablecaption{Properties of photometric systems and point source catalogs}
\startdata
\tablehead{Instrument & Epoch\tablenotemark{a} & Filter name & $\lambda$ & magnitude limits & $\sigma_{\rm calibration}$\tablenotemark{b} \\
& [year] & & [$\mu$m] & [mag] &}
  {\it Gaia} eDR3     & 2016 & $G$, $G_{\rm BP}$, $G_{\rm RP}$ & 0.622, 0.511, 0.777\tablenotemark{piv} & m$_{G}$: --21 & 1 $\%$ \\
  WISE          & 2010.5 & $W1$, $W2$ & 3.353, 4.603\tablenotemark{iso} & 2--17.1, 1.5--15.7 & 1.5, 1.5 $\%$ \\
                & & $W3$, $W4$ & 11.561, 22.088 & -3--11.4, -4--8.0 & 1.5, 3.5 \\
  SEIP          & 2006 & $I1$--4 & 3.550, 4.493, 5.731, 7.872 & & 3 $\%$ \\
                & & $M1$ & 23.68 & & 2--3 $\%$ \\
\hline
  PanSTARRS DR1 & 2012.5 & $g_{\rm P1}, \,r_{\rm P1}, \,i_{\rm P1}$ & 0.481, 0.617, 0.752\tablenotemark{piv} & 13.5--19.1 (AB) & 0.02 mag \\
                & & $z_{\rm P1}, \,y_{\rm P1}$ & 0.866, 0.962 & 13.0--, 12.0-- \\
  SkyMapper DR1.1 & 2015 & $u,\, v$ & 0.349, 0.384\tablenotemark{cen} & 8.5--22 (AB) & 1 $\%$ \\
                & & $g,\, r,\, i,\, z,$ & 0.510, 0.617, 0.779, 0.916 \\
  APASS DR9     & 2011 & $B,\, V$& 0.445, 0.551\tablenotemark{c} & m$_{V}$: 7.5--17 & 0.013, 0.012 mag \\
                & & $g',\, r',\, i'$& 0.483, 0.626, 0.767 & & 0.012, 0.014, 0.021 \\
  2MASS         & 1999 & $J$, $H$, $K_{\rm s}$                    & 1.235, 1.662, 2.159\tablenotemark{iso} & --15.8, --15.1, --14.3 & 1.7, 2.0, 1.9 $\%$
\enddata
\tablecomments{
  (piv) Pivot wavelengths ; (iso) Isophotal wavelength ; (cen) Central wavelength.
  (a) The mean epochs are obtained by averaging the observational periods of the missions, except for {\it Gaia} eDR3. 
  (b) Systematic uncertainties including calibration errors.
  (c) These wavelengths, obtained from \cite{fukugita96}, are presented as references for the APASS DR9. \\
  The following references are provided for the items of each mission: {\it Gaia} eDR3: \cite{riello21}; WISE: \cite{wise}; SEIP: \cite{seip}; PanSTARRS DR1: \cite{tonry12} and \cite{magnier13}; SkyMapper DR1.1: \cite{wolf18}; APASS DR9: \cite{apass} and \cite{munari14}; and 2MASS: \cite{skrutskie06}.
  }
\label{tab1}
\end{deluxetable*}


\begin{figure}[htbp]
  \epsscale{1.15}
  \plotone{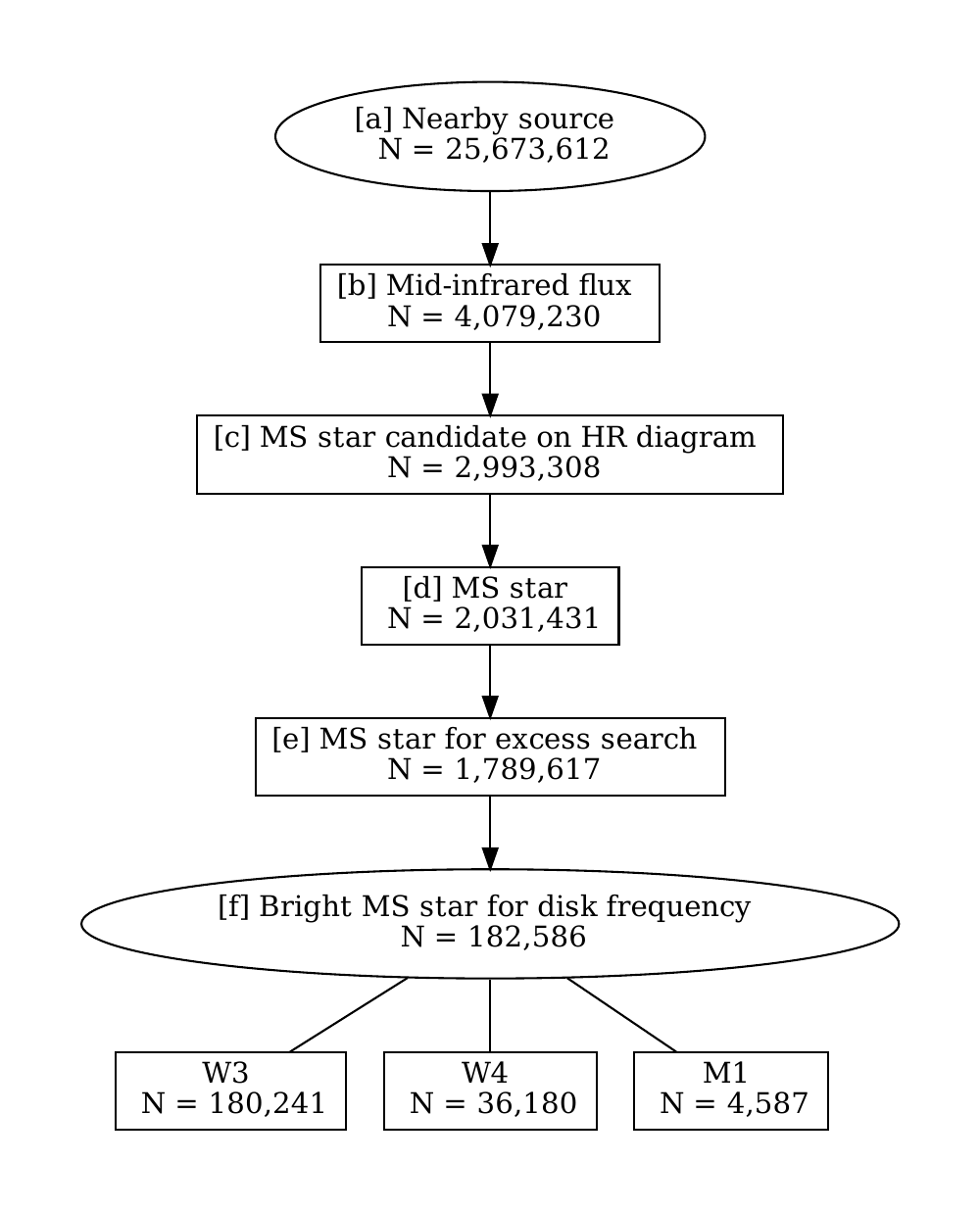}
  \caption{
  Sample selection and the number of stars reduced by each criterion are shown as a flowchart: 
  [a] nearby sources selected from {\it Gaia} archive (Section \ref{sec:gaiastar}), 
  [b] sources with mid-infrared fluxes (Section \ref{sec:midflux}), 
  [c] sources within the defined regions of main sequence stars (Section \ref{sec:cmd}), 
  [d] main sequence stars with the effective temperature of 3000--15000 K (Section \ref{sec:remove}), 
  [e] main sequence stars whose photospheric fluxes can be detected with 3$\sigma$ confidence levels to the detection limits at mid-infrared wavelength (Section \ref{sec:identifyexcess}), and 
  [f] stars that are bright enough to avoid the contamination of background objects (Section \ref{sec:contami}).
  }
  \label{fig:flowchart}
\end{figure}

\section{Selecting nearby sources} \label{sec:data}
An infrared excess is the flux excess in the infrared bands relative to the stellar SED estimated from the shorter-wavelength bands.
Thus, to search for infrared excess, we must first compile photometric data for the optical as well as near-infrared wavelengths.
Table \ref{tab1} summarizes the catalogs used in this work, which are explained in detail in the following subsections. 
Figure \ref{fig:flowchart} summarizes how stars with infrared excess were selected in our analyses in Sections \ref{sec:data}--\ref{sec:correction}.

\subsection{Stars with known parallax} \label{sec:gaiastar}
To build the sample, we began with the {\it Gaia} catalog \citep{gaia}. 
It contains parallaxes and magnitudes at three wavelength bands $G$, $G_{\rm BP}$, and $G_{\rm RP}$ (Table \ref{tab1}).
In the early data release 3 (eDR3), the detection limit reached about 21 mag in the $G$-band, and about 0.6 billion sources with five-parameter astrometry are contained in the catalog.
Typical astrometric precision is in the range of 0.02--0.03 and 0.5 mas for stars with m$_{G}$ = 9--14 and 20 mag, respectively \citep{gaiaedr3}. \par

We selected sources from the {\it Gaia} archive using the following criteria:
To avoid significant background contamination, sources in the crowded galactic plane ($|b|<10^{\circ}$) and in regions close to large magellanic cloud or nearby galaxies, were excluded.
To select {\it Gaia} sources whose photospheric emission is detectable in mid-infrared bands discussed in this work, we limited the parallax of the target stars to intermediate-mass stars ($G_{\rm BP}-G_{\rm RP} \le 0.5$) and low-mass stars  ($G_{\rm BP}-G_{\rm RP} > 0.5$) within distances of $\sim$ 1 kpc and $\sim$ 500 pc, respectively.
As the first step in selecting a sample to search for the infrared excess, we selected $\ngaia$ sources ([a] in Figure \ref{fig:flowchart}). \par

This work discusses the frequency of debris disk emission in three mid-infrared catalog data products, WISE/$W3$, $W4$, and {\it Spitzer}/MIPS ch1. 
The above distance thresholds have been tentatively set to fully cover {\it Gaia} sources detectable with the sensitivity of $W4$ \citep[6 mJy,][]{wise}, which has the lowest sensitivity to stellar photospheric emission. However, distant sources that are not bright enough to detect their photospheric emission in $W4$ are excluded after SED fitting (based on Equation (\ref{eq:photsphere}) in Section \ref{sec:sed}). On the other hand, some sources detectable with $W3$ and MIPS ch1 are missed by these distance thresholds. After the final correction for contamination by non-debris emission, it is shown that very few debris disks are missed in all bands, indicating that the above distance thresholds do not affect completeness at the current sensitivity limits (see Section \ref{sec:contami} for more details). \par

\begin{figure*}[htbp]
  \epsscale{1.15}
  \plotone{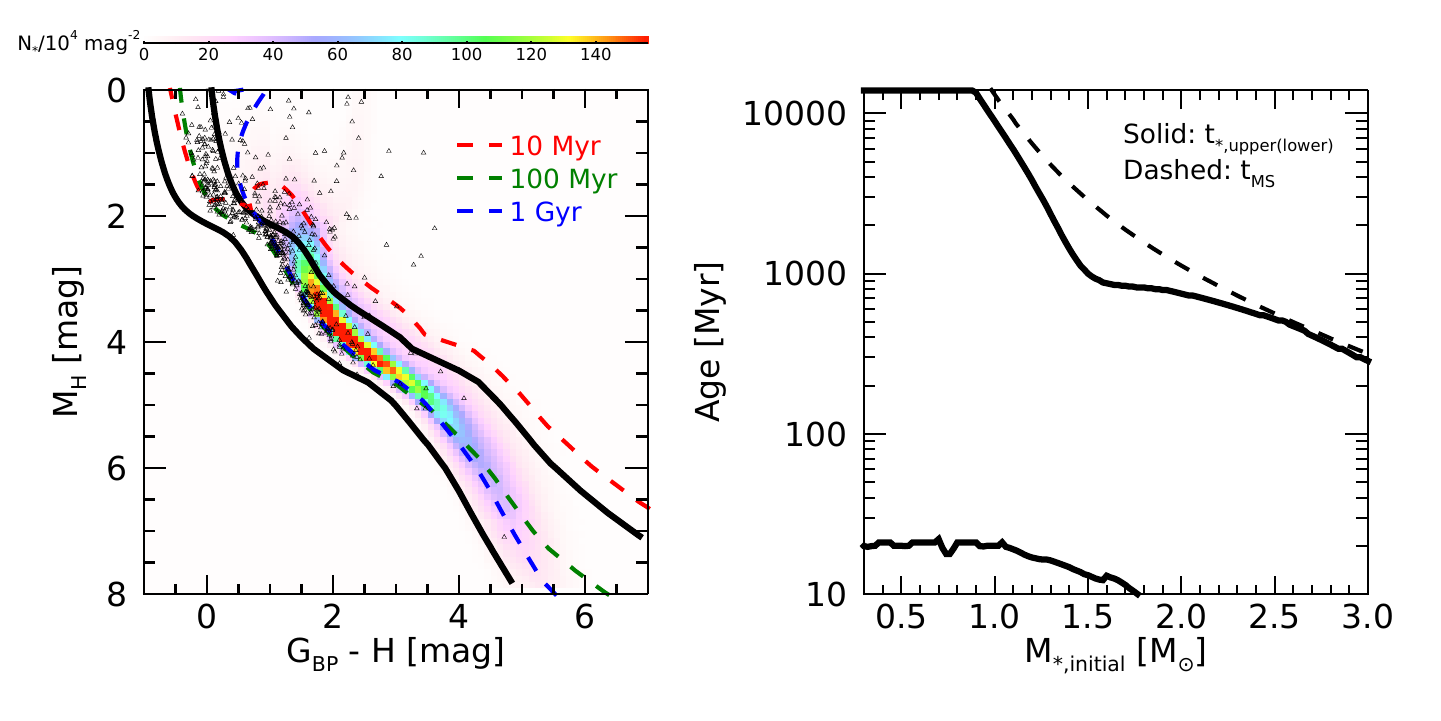}
  \caption{
  The left panel shows the distribution of the sample where infrared excess was searched for in the color-magnitude diagram.
  The number density of stars is shown as a two-dimensional histogram in color.
  The triangles indicate the stars hosting debris disks that have already been detected and identified as 'Prime' infrared excess stars in \citep{cotten16}.
  Dashed lines show PARSEC V1.2S isochrones with the ages of 10 Myr, 100 Myr, and 1 Gyr \citep{bressan12}. 
  In this work, the location where main sequence stars are present is defined by thick black lines. 
  The right panel shows the possible range of isochronal ages ($t_{\rm *,upper(lower)}$) for the sample as a function of stellar mass, where the dashed line represents the classical timescale of main sequence phases as $t_{\rm MS*} \sim t_{\odot} (M_{*}/M_{\odot})(L_{\odot}/L_{*}$) and $t_{\odot} \sim$ 10 Gyr.
  }
  \label{fig:cmd}
\end{figure*}

\subsection{Mid-infrared photometry to identify excess} \label{sec:midflux}
We used the infrared data obtained by WISE and {\it Spitzer}.
The WISE satellite observed almost the entire sky in the cryogenic phase \citep{wise}.
In contrast, {\it Spitzer} achieved higher sensitivity and resolution \citep{spitzer}, although its coverage of the sky is limited.
Assuming that the sensitivities of WISE/$W4$ and {\it Spitzer}/MIPS ch1 are 6 mJy and 0.1\footnote{about 12 mag, which is under optimal conditions with integration time of 500 seconds \citep{rieke04}.} mJy, respectively, sun-like stars within the distances of about 80 and 500 pc should be detected by those missions.
The infrared archival data are available from the IRSA of NASA/IPAC. 
We downloaded parts of the source catalogs of AllWISE \citep[][DOI: 10.26131/IRSA1]{cutri12} and SEIP \citep[{\it Spitzer} Enhanced Imaging Product,][DOI: 10.26131/IRSA3]{seip} and cross-identified them with the nearby sources.
Nearby stars have large proper motions that can cause cross-identifications to fail. 
{\it Gaia}'s coordinates were traced back to the mean epochs of each mission using the proper motions, and we re-identified a list of nearby mid-infrared sources. 
The WISE and SEIP catalogs (whose spatial resolution is poorer than other auxiliary catalogs) have astrometric uncertainties smaller than 0.5" and 0.3", respectively, which are given on a website\footnote{\url{https://wise2.ipac.caltech.edu/docs/release/allsky/expsup/sec6_4.html}} and in Section 4.8 of the SEIP Explanatory Supplement\footnote{\url{https://irsa.ipac.caltech.edu/data/SPITZER/Enhanced/SEIP/overview.html}}.
To account for possible positional differences due to parallax and a single-valued epoch for each mission, we simply use a radius of 1 arcsec for cross-identification between all the catalogs in this work.
As a result, $\nmir$ {\it Gaia} sources remained ([b] in Figure \ref{fig:flowchart}), which had some of the mid-infrared photometric data at 12, 22, and 24 $\mu$m for WISE/$W3$, WISE/$W4$, and SEIP/$M1$, respectively.

\subsection{Photometric data to characterize host stars}
Because {\it Gaia} performs photometry in only three wavelength bands that are close to each other, we compiled additional photometric data to build a more accurate stellar SED to search for the infrared excess. 
Table \ref{tab1} summarizes all the catalogs used in this work: Panoramic Survey Telescope and Rapid Response System \citep[PanSTARRS DR1,][]{chambers16}, SkyMapper DR1.1 \citep{keller07, wolf18}, AAVSO Photometric All Sky Survey \citep[APASS DR9,][]{apass, munari14}, Two Micron All Sky Survey \citep[2MASS,][]{skrutskie06}, {\it Spitzer,} WISE, and {\it Gaia}.
We cross-identified a list of nearby mid-infrared sources, and downloaded parts of the catalogs from Centre de Donn\'ees astronomiques de Strasbourg (CDS\footnote{\url{http://cdsxmatch.u-strasbg.fr/}}). \par

\subsection{The main sequence star candidates} \label{sec:cmd}
To focus on the debris disks around main sequence (MS) stars, we define a region where main sequence stars dominate in color-magnitude diagrams (the thick black lines in Figure \ref{fig:cmd}).
Due to the strict definition of the MS region, some of the debris disks that are slightly off the main sequence in previous studies \citep[e.g.,][]{cotten16}, are not included in our sample.
The MS region is obtained via the following steps.
First, we set the reference line of the main sequence regions by combining two isochrones: 100 Myr for $M_{*,{\rm initial}} > 0.8 M_{\odot}$ and 1 Gyr for $M_{*,{\rm initial}} < 0.8 M_{\odot}$, considering the difference in main sequence timescales on the stellar masses. 
For this, we employed the theoretical isochrones of PAdova and TRieste Stellar Evolution Code V1.2S \citep[PARSEC,][]{bressan12}.
Second, we shift the reference line horizontally by 0.5 mag in the negative direction, defining a blue limit for the MS region (the left thick black line in the left panel of Figure \ref{fig:cmd}).
Finally, the red limit is determined by combining two curves: one with the reference line horizontally shifted by 0.5 mag in the positive direction for $M_{*,{\rm initial}} > 1.1 M_{\odot}$, and the isochrone of 20 Myr for $M_{*,{\rm initial}} \lesssim 1.1 M_{\odot}$. 
This approach helps prevent the exclusion of young low-mass stars.
We performed isochrone fitting on the color-magnitude locations surrounded by the MS region. 
The upper and lower limits for stellar ages ($t_{\rm *,upper(lower)}$) as a function of stellar mass are obtained for the MS star candidates (the right panel of Figure \ref{fig:cmd}). \par

We used optical fluxes from {\it Gaia} ($G, \ G_{\rm BP}$) and near-infrared fluxes from 2MASS ($J, \ H, \ K_{\rm s}$) or WISE ($W1$) to create eight independent color-magnitude diagrams, where the color between the optical/near-infrared bands is plotted against the near-infrared absolute magnitude.
According to their positions in these diagrams, the stars are classified into five categories; (1) stars in the MS region, (2) stars bluer than the region, (3) stars redder than the region, (4) stars out of the diagrams, and (5) stars with the invalid color or magnitude due to missing or erroneous observations. 
Due to photometric uncertainties, the eight color-magnitude diagrams may yield multiple classifications for a star, especially near the thresholds of the MS region.
In such a case, the most frequent classification obtained after excluding the cases classified as (5) is used.
As a result, we found $\ncmd$ MS star candidates ([c] in Figure \ref{fig:flowchart}).

\begin{figure*}[htbp]
  \epsscale{1.05}
  \plotone{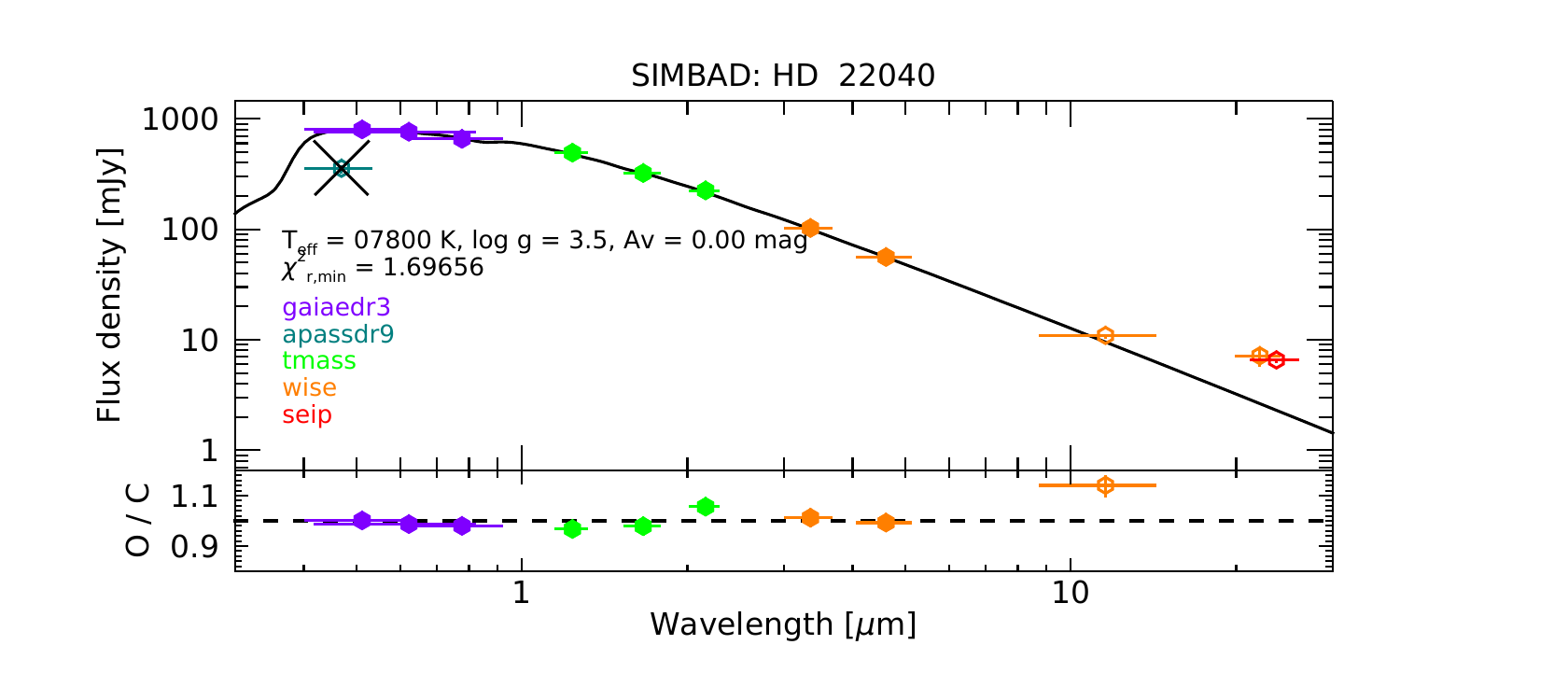}
  \caption{
  An example of a star with infrared excess according to our analysis, in which only photometric data with filled symbols are fitted to determine stellar parameters.
  The crossed-out symbol indicates the data ignored by the iteration in the fitting (Section \ref{sec:sed}).
  The bottom panel shows the ratio between the observed and calculated fluxes for stellar radiation.
  }
  \label{fig:sed}
\end{figure*}

\begin{figure}[htbp]
  \epsscale{1.15}
  \plotone{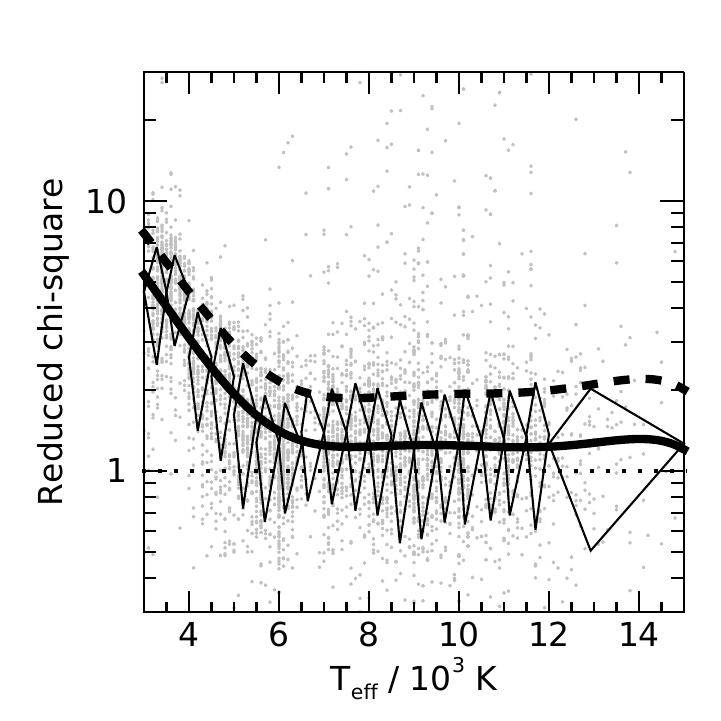}
  \caption{
  Reduced chi-square obtained from SED fitting for $\nchisq$ stars as a function of the stellar effective temperature.
  The diamonds indicate the effective temperature bins with the predefined size of 500 K and the typical 1$\sigma$ scatter of the reduced chi-square, for the  horizontal and vertical extent, respectively. 
  If the number of stars in a bin is less than 100, the bin is merged with its neighbors until the number exceeds the threshold.
  The solid and dashed lines show the best-fit solution with a polynomial function and the acceptable ranges of the 1$\sigma$ confidence level for the sample candidates in Section \ref{sec:remove}.
  }
  \label{fig:chisq}
\end{figure}

\begin{figure*}[htbp]
  \centering
  \epsscale{1.15}
  \plotone{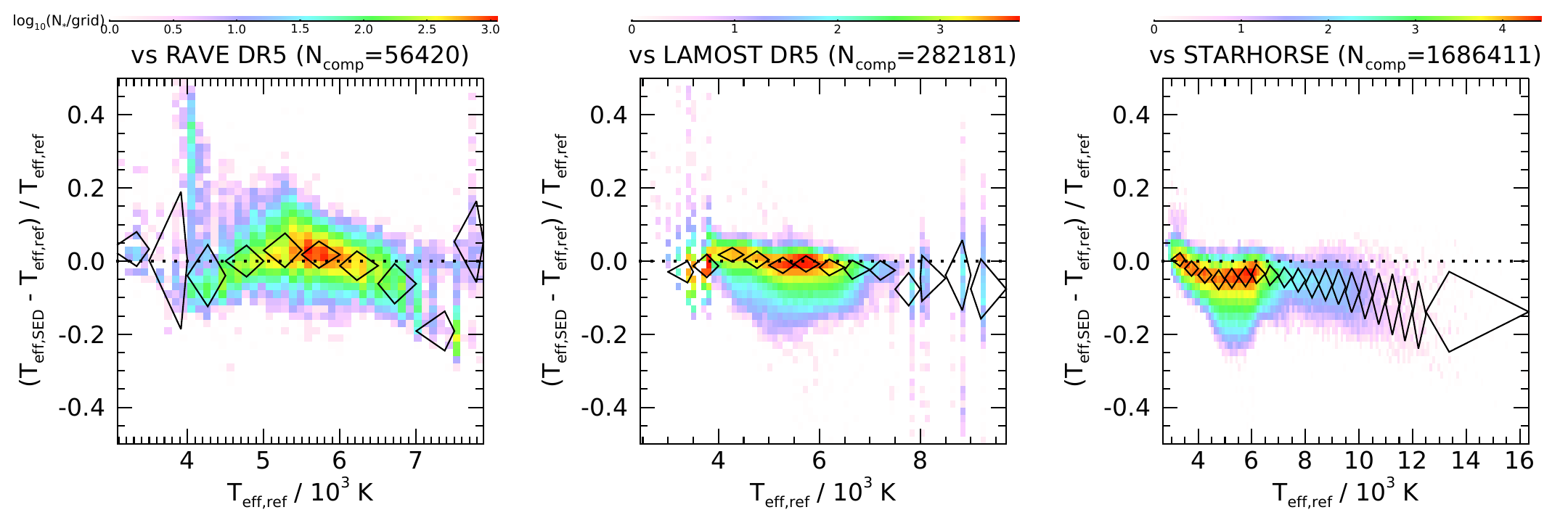}
  \caption{
  The stellar effective temperatures are compared with those in RAVE DR5, LAMOST DR5, and StarHorse from left to right.
  The y-axis indicates the fractional difference of the effective temperatures obtained by the SED fitting in this work and those of previous studies.
  The diamonds indicate the effective temperature bins with the predefined size of 500 K and the typical 1$\sigma$ scatter of the reduced chi-square, for the  horizontal and vertical extent, respectively. 
  If the number of stars in a bin is less than 100, the bin is merged with its neighbors until the number exceeds the threshold.
  }
  \label{fig:dteff}
\end{figure*}

\section{Analysis of the spectral energy distribution} \label{sec:analysis}

\subsection{Flags and cautions in catalogs} \label{sec:flags}
The catalogs contain flags that indicate the reliability of the data: position, flux, and the presence of the source itself.
By checking the flags and cautions described in the references, we carefully removed the suspicious data: low quality data (for some given reason), extended objects, variable fluxes, and fluxes beyond the saturation limits. 
In the case where the saturation limits are relaxed, such as by observations with short exposure times and fitting to wings of the point spread function (PSF), we use the relaxed saturation limits.
We do not use upper limits for fluxes.
Possible systematic errors were also examined and propagated to the total error budget.
The procedure is as follows:
\begin{itemize}
\item {\it Gaia} eDR3: The photometric system of {\it Gaia} has calibration errors of 2.0, 3.1, and 1.8 milli-mag (mmag) for $G, \ G_{\rm BP}$, and $G_{\rm RP}$, respectively, and the overall accuracy of the calibration is about 1 $\%$ \citep{riello21}. 

\item PanSTARRS DR1: 
The binary-encoded flags 'qualityFlag' (renamed as 'Qual' on the CDS servers) and `$\lambda$Flags' ($\lambda$ stands for each filter name) are used to select reliable data.
The Qual flags indicate whether an object is real or a likely false positive, and their 64 and 128 (6 and 7 for the binary-encoded flag) are particularly relevant for suspect or low-quality objects \citep[see datails for the flags,][]{flewelling20}.
We removed objects identified by the two in the Qual flags.
The $\lambda$flags are relevant to the detection of objects at each filter.
Objects identified by the Qual flag of 1 and 2 (0 and 1) or the $\lambda$flags of 8192 (13) are removed as extended objects potentially caused by contamination.
\cite{tonry12} reported systematic errors of 0.02 mag in the photometric calibration. 

\item SkyMapper DR1.1: 
A binary-encoded flag `flags' is used, and objects identified by any of them are removed.
\cite{wolf18} determined that the reproducibility of the photometry was about 1 $\%$.
We found that the $u$- and $v$-band fluxes tended to disagree with the models in the SED fitting, and therefore decided not to use these data. 
While the reason for the discrepancy at shorter wavelengths is beyond the scope of this work, ignoring some parameters such as metallicity may cause the result (see a later Section \ref{sec:sed}).

\item APASS DR9: 
The internal accuracies vary from 0.012 to 0.021 mag for different filters \citep{munari14}.
Similar to the case of the $u$- and $v$-band fluxes in SkyMapper DR1.1, we decided not to use the $B$-band fluxes for the SED fitting.

\item 2MASS:
The 'ph$\_$qual' flag (Qflg) indicates the photometric quality of objects at each filter (see datails for the 2MASS flags on a website\footnote{\url{https://irsa.ipac.caltech.edu/data/2MASS/docs/releases/allsky/doc/sec2_2a.html}}).
While A--D of the flag are relevant for the signal-to-noise ratio of the detection, others of the flag indicate particular cases such as U for the upper limits of the fluxes.
The 'gal$\_$contam' flag (Xflg) is relevant for extended sources, and objects with Xflg = 0 are free of significant contamination.
We use only objects and fluxes with Qflg = A--D and Xflg = 0 to characterize the stellar properties.

\item AllWISE:
The concepts of the `qph' and `ext$\_$flg (ex)' flags are similar to the 2MASS ph$\_$qual and gal$\_$contam flags, although the definitions are different (see details for the WISE flags\footnote{\url{https://wise2.ipac.caltech.edu/docs/release/allwise/expsup/sec2_1a.html}}).
The `cc$\_$flags (ccf)' flag is relevant to image artifacts, and sources with ccf = 0 are not affected by known artifacts.
The `var$\_$flg (var)' flag indicates the variability of objects, and sources with var = 0--5 are most likely not variable.
In summary, we use only objects and fluxes with qph = A--C, ex = 0, ccf = 0, var $<$ 6 to characterize stellar properties.

\item SEIP: To select flux data with better accuracy, we applied several criteria as recommended in Section 1.1.2 of SEIP Explanatory Supplement. The calibrator fluxes obtained by two approaches are reported to be consistent within an rms error of a few percent in Section 4.1.2 in MIPS Instrument Handbook\footnote{\url{https://irsa.ipac.caltech.edu/data/SPITZER/docs/mips/mipsinstrumenthandbook/}}. We adopt the systematic uncertainty of 3$\%$ for all the photometry of SEIP.
\end{itemize}

\subsection{SED fitting for stellar properties} \label{sec:sed}
To characterize the stellar properties, we fitted broadband photometric fluxes with reference wavelengths shorter than 5.0 $\mu$m using stellar atmospheric models.
We adopted the stellar atmospheric models of BT-Settle and BT-Nextgen for stars with effective temperatures of 3000--7000 K and 7200--20000 K, respectively \citep{allard12}.
The precomputed grids of the models are available from the Spanish Virtual Observatory (SVO\footnote{\url{http://svo2.cab.inta-csic.es/theory/newov2/index.php?models=bt-settl-cifist}}), 
but the grid sizes for early-type stars are not sufficiently small for our purposes: 100 K for the BT-settle models, 200 K for 7000--12000 K, and 500 K for 12000--20000 K. 
Therefore, we obtained data with a grid of about 100--200 K by interpolating the values in the original models.
To compare observed broadband fluxes with model SEDs, a color correction must be performed. 
Instead of applying a single-valued color correction factor of $K$ to the observed fluxes \citep[e.g., Section V\hspace{-1.2pt}I C.3 of][]{beichman88}, we convolved models with filter response functions to match the photometric systems of the observations \citep[see for example, Appendix A of][]{robitaille07}.
We first obtained a rough estimate of a stellar parameter from the {\it Gaia} fluxes and fitted the stellar models to all available data using least-square minimization.
The fitting was performed iteratively, and outlier fluxes were removed.
During the fitting, we changed the stellar effective temperature and surface gravity, while fixing [Fe/H] = 0 and $A_{V}$ = 0. 
Although these parameters, fixed in the fitting, can change the shape of the SED, especially at shorter optical wavelengths, the identification of infrared excess at mid-infrared wavelengths is not affected. \par

The observed photometric data are consistent with the model SEDs, particularly for intermediate-mass stars (Figure \ref{fig:sed}).
The reduced chi-square of the SED fitting shows a clear dependence on the stellar temperature; while the chi-square value is typically 1.3 for early K- to late B-type stars, the chi-square for M dwarfs is up to approximately 5 (Figure \ref{fig:chisq}).
We did not investigate the details of this inconsistency because it was beyond the scope of this work.
Briefly, late-type stars have more molecular lines in their atmosphere, which are difficult to be reproduced by theoretical models.
In addition, the fixed metallicity in SED fitting may cause discrepancies at shorter wavelengths.
A similar tendency is recognized in Sections 3.1 and 4.2 of \cite{casagrande14}, although they used different datasets and models. 
We do not evaluate the excess emission to photospheres in terms of absolute values, but in terms of relative deviations from the mean (see later Section \ref{sec:identifyexcess}).
Therefore, the poor chi-squared fitting has a negligible effect on the identification of the infrared excess.
\par
 
To check the robustness of the current analysis, Figure \ref{fig:dteff} shows comparisons of the derived effective temperatures with those obtained in 3 large stellar surveys: the Radial Velocity Experiment \citep[RAVE,][]{kunder17}, the Large Sky Area Multi-Object Fibre Spectroscopic Telescope \citep[LAMOST,][]{luo12, zhao12}, and StarHorse \citep{queiroz18, starhorse2021}.
The RAVE is a spectroscopic survey of stars with a medium resolution of R $\sim$ 7500.
LAMOST performs spectroscopic surveys with lower spectral resolutions of R = 500 and 1800, enabling the characterization of fainter stars. 
StarHorse is a Bayesian inference code that determines stellar parameters.
Using several photometric catalogs and the parallax measured by {\it Gaia}, it provides a catalog of stellar parameters.
As shown in Figure \ref{fig:dteff}, our estimates are consistent with those of previous studies.
Although we identified small systematic deviations ($\Delta T_{\rm eff}/T_{\rm eff} \sim 0.1$), these differences are unlikely to affect the identification of the infrared excess.

\subsection{Removal of outliers} \label{sec:remove}
We enforced a few target selection criteria based on the SED fitting results.
In a binary system with a late-type companion, the companion is very faint at optical wavelengths but can cause a weak excess from the primary emission in the near- and mid-infrared with a small wavelength dependence (see Appendix \ref{sec:app} for more details).
Therefore, from the sample candidates, we excluded the stars whose averaged deviation from the model between 2 and 5 $\mu$m was greater than the 3$\sigma$ confidence.
Imposing this condition makes the detection of disks with $T_{\rm disk} > 400$ K difficult (see Section \ref{sec:diffband} for details). \par

Binary candidates are also identified using the {\it Gaia} catalog.
The inconsistency between the {\it Gaia} fluxes $G$, $G_{\rm BP}$ and $G_{\rm RP}$ 
is provided as $C$ (phot$\_$bp$\_$rp$\_$excess$\_$factor) in the {\it Gaia} catalog.
The factor $C$ has a color dependence that can be corrected using Equation (6) and Table 2 of \cite{riello21}, and the corrected factor is given as $C^{*}$.
The typical scatter of $C^{*}$ as a function of $G$ magnitude, $\sigma_{C^{*}}(G)$, is given by Equation (18) of \cite{riello21}.
To identify binary candidates, we did not use the deviation in the fitting with the astrometric solution, although the deviation (`astrometric$\_$excess$\_$noise$\_$sig') can be significant for binaries \citep[Section 5.3 of][]{lindegren21}.
This is because almost all intermediate-mass stars will be removed if this criterion is adopted.
We do not pursue the reason for this, but the higher binary rate for massive stars may be possible.
In summary, we used only the criterion for the photometric inconsistency in the {\it Gaia} catalog and removed $\gaiabinary$ stars with $C^{*}/\sigma_{C^{*}}(G) > 2$ as possible binaries. \par

Furthermore, we excluded from the sample candidates those sources that met any of the following criteria: 
\begin{itemize}
    \item Outlier fluxes were automatically ignored in the SED fitting via iteration (shown as a cross in Figure \ref{fig:sed}). 
    If the fraction of rejected data points is greater than 0.5, those stars were excluded from the sample candidates.
    \item A possible hot star with a stellar effective temperature $>$ 15000 K was excluded because the number of such objects was too small for any statistical discussion.
    \item A possible brown dwarf with a stellar effective temperature $<$ 3000 K was excluded because the stellar mass could not be determined from its effective temperature. Due to the long-term evolution without hydrogen burning at the core, the effective temperature of a brown dwarf is degenerated between the mass and age. Furthermore, its SEDs has a complicated structure due to molecular absorption, which is difficult to accurately characterize using only photometric fluxes.
\end{itemize}

To obtain the reference stellar flux for the infrared excess, the SEDs of host stars must be determined accurately.
Owing to the incompleteness of the adopted models in the fitting, the average reduced chi-square for each narrow range of the stellar effective temperature did not converge to one.
Therefore, we expressed the average reduced chi-square and its uncertainty using polynomial functions of the stellar effective temperature (Figure \ref{fig:chisq}).
Stars with chi-square values outside this range, shown in Figure \ref{fig:chisq}, were removed from the sample candidates.
As a result, $\nms$ main sequence stars remained as sample candidates ([d] in Figure \ref{fig:flowchart}). \par

\subsection{Removing targets with faint mid-infrared fluxes}
In sensitivity-limited surveys of infrared excess, some faint stars with photospheric fluxes below the detection limits in the relevant catalogs can be detected owing to the presence of a large infrared excess.
Without removing such stars from the sample, the derived frequencies of the debris disks are overestimated, as noted by \cite{fujiwara13}.
To mitigate this bias, we search for infrared excess at each wavelength by including only stars that meet the criterion for their photospheric flux, extrapolated by SED fitting, $f_{\rm *,model}$, as follows:
\begin{eqnarray}
  f_{\rm *,model} > 3 \sigma_{\rm obs}, 
  \label{eq:photsphere}
\end{eqnarray}
where $\sigma_{\rm obs}$ is the observational uncertainty.
If the criterion was not applied, an overestimation of the disk frequencies of low-mass stars by up to a few orders was found at $W4$. 
Therefore, to search for the infrared excess, only $\nphot$ stars remained in the sample ([e] in Figure \ref{fig:flowchart}).

\subsection{Identification of infrared excess} \label{sec:identifyexcess}
\begin{figure*}[htbp]
  \epsscale{1.15}
  \plotone{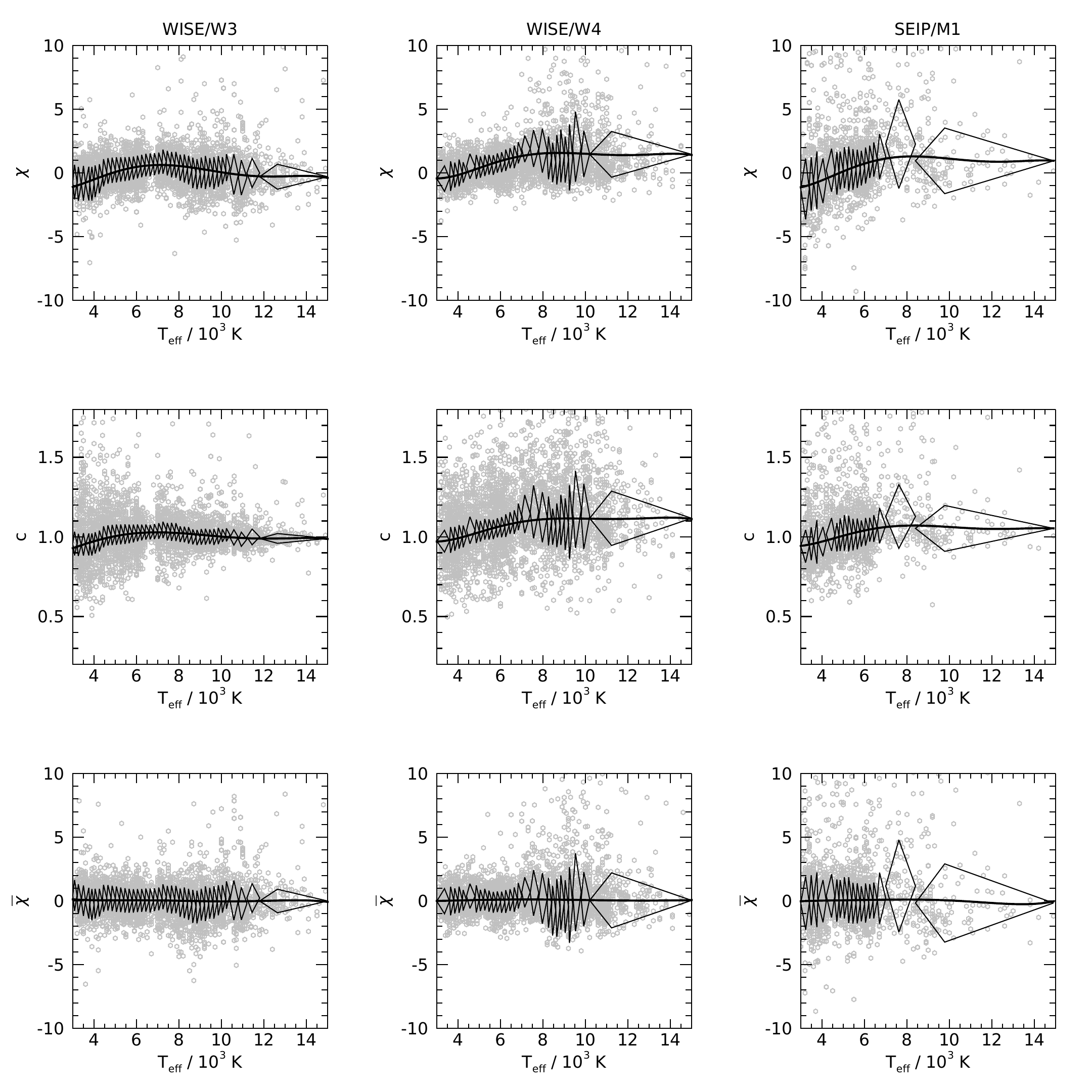}
  \caption{
  The significance of the excess, the correction coefficient to match the models to the observed fluxes using the average excess in each $T_{\rm eff}$ bin, and the significance of the corrected excess represented as $\chi$, $c$, and $\bar{\chi}$ in Equations (\ref{eq:soe}), (\ref{eq:c}) and (\ref{eq:chibar}), respectively, are shown as functions of the stellar effective temperature.
  The diamonds indicate the effective temperature bins with the predefined size of 200 K and the typical 1$\sigma$ scatter of the reduced chi-square, for the  horizontal and vertical extent, respectively. If the number of stars in a bin is less than 100, the bin is merged with its neighbors until the number exceeds the threshold.
  The best-fit polynomial solutions for the diamonds are also shown.
  }
  \label{fig:chibar}
\end{figure*}

The significance of the excess can be represented by $\chi$ as follows:
\begin{eqnarray}
  \chi &=& \frac{f_{\rm obs}-f_{\rm *,model}}{\sigma_{\rm obs}}
  \label{eq:soe}
\end{eqnarray}
where $f_{\rm obs}$ denotes the observed flux.
Because most objects have no significant excess from stellar radiation, the histogram of $\chi$ should ideally peak at zero, with a small addition of positive values owing to infrared radiation from debris disks. 
However, we identified significant nonzero offsets for the average numbers, as shown in the top panels of Figure \ref{fig:chibar}.
The averaged $\chi$ is negative for low-mass stars, but positive for stars with higher temperatures at the three wavelengths.
The reason for this systematic trend is unclear; however, it could be due to incomplete stellar models or biases from the assumption of fixed solar metallicity and no dust extinction in the current analysis.
To compensate for this, we corrected the model fluxes so that the average ratio between the observed and model fluxes, 
\begin{eqnarray}
  c &=& \frac{f_{\rm obs}}{f_{\rm *,model}}
  \label{eq:c}
\end{eqnarray}
was one.
The middle panels of Figure \ref{fig:chibar} show the distribution of $c$ as a function of the stellar temperature.
We fitted $c$ with polynomial functions to obtain its typical value as a function of the stellar temperature, denoted by $\langle c \rangle$. 
The significance of the infrared excess is then defined as follows:
\begin{eqnarray}
  \bar{\chi} &=& \frac{f_{\rm obs}-\langle c \rangle f_{\rm *,model}}{\sigma_{\rm obs}}.
  \label{eq:chibar}
\end{eqnarray}
The bottom panels of Figure \ref{fig:chibar} show $\bar{\chi}$ as a function of the effective temperatures of the host stars.
Using 3$\sigma_{\rm obs}$ confidence ($\bar{\chi} \geq$ 3) to identify infrared excesses, $\newiset$, $\newisef$, and $\neseipm$ stars were found to exhibit infrared excesses for $W3$, $W4$, and $M1$, respectively (Table \ref{tab2}).

\begin{deluxetable}{ccc ccc} \label{tab2}
\tablecaption{The numbers of stars and those with excess emission.}
\startdata
\tablehead{Band & $N_{\rm *}$ & $N_{\rm excess,raw}$}
W3 & 1787916 & 15166 \\
W4 & 36180 & 583 \\
M1 & 7141 & 575 \\
\enddata
\end{deluxetable}

\section{Problems in deriving the frequency of debris disks using infrared excess} \label{sec:correction}
A star can show infrared excess not only due to a debris disk but also due to other factors such as contamination from background objects.
These factors were almost ignored in the calibration-limited surveys of previous studies, which identified only bright infrared excesses.
By contrast, the sample for infrared excess constructed in Sections \ref{sec:data}--\ref{sec:analysis} is based on sensitivity-limited surveys. 
The large number of stars is helpful for discussing the central stellar-type dependence of the debris disks. 
However, the detection limits of the infrared excess in terms of excess flux over stellar flux vary from source to source in a sensitivity-limited survey, which must be corrected to estimate a reliable debris disk frequency under a uniform criterion, as described in Section \ref{sec:efficiency}.
In addition, stars for which background contamination cannot be ruled out, even if infrared excesses are detected, should be excluded from the sample. 
This is discussed in Section \ref{sec:contami}.

\begin{figure*}[htbp]
  \epsscale{1.15}
  \plotone{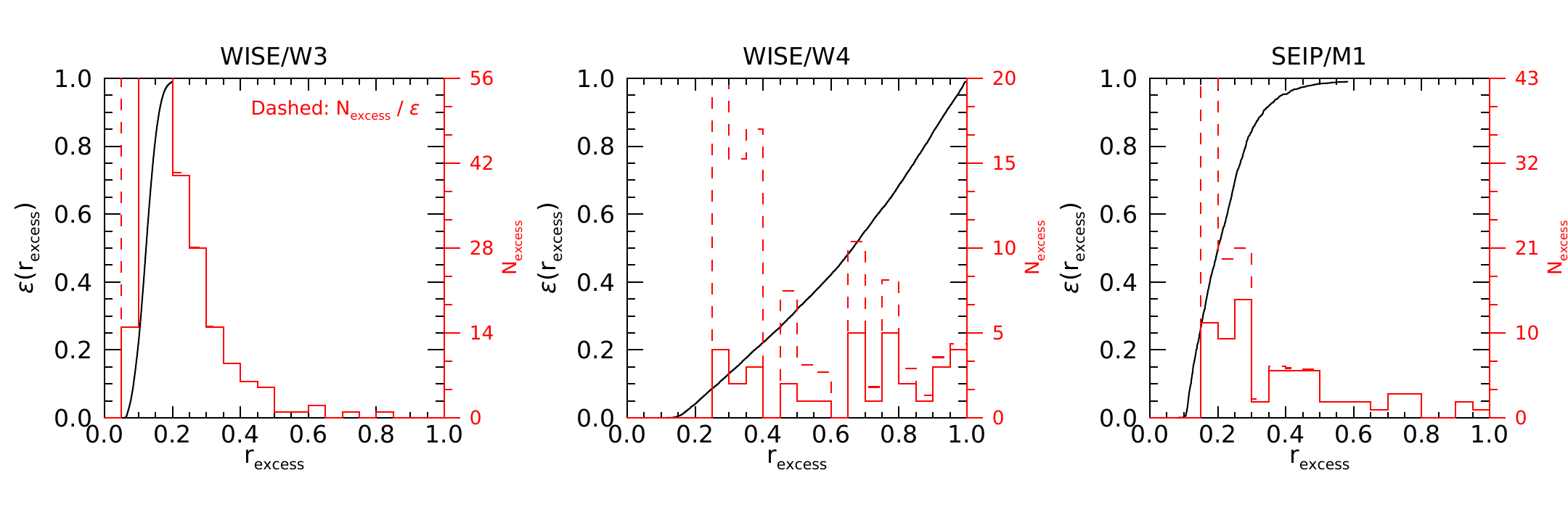}
  \caption{
  Detection efficiency of infrared excess for stars whose effective temperatures are between 5000--6000 K is shown for $W3$, $W4$, and $M1$ from left to right. 
  The number of infrared excesses as a function of excess ratio is also presented as histograms with the right y-axis colored in red.
  }
  \label{fig:efficiency}
\end{figure*}

\begin{figure*}[htbp]
  \epsscale{1.15}
  \plotone{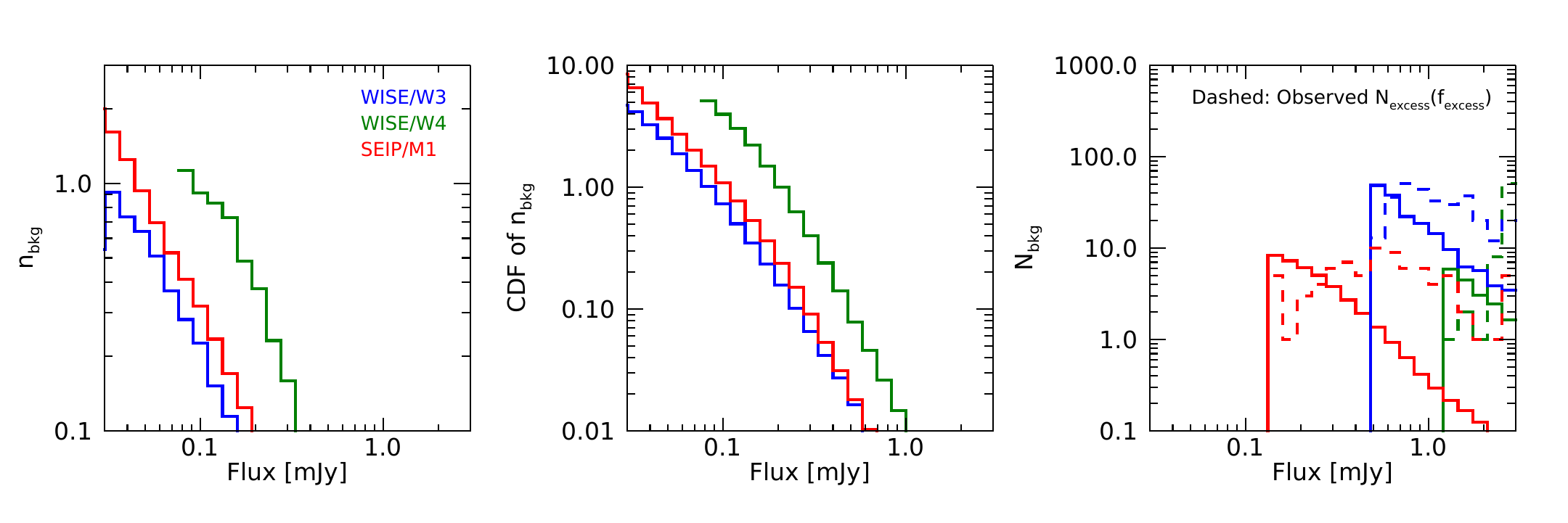}
  \caption{
  \textit{(left)} The number of background sources within an aperture based on Equation (\ref{eq:bkg1}) for $W3$, $W4$, and $M1$.
  The number densities of WISE are based on {\it AKARI} observations (see Section \ref{sec:contami}). 
  \textit{(middle)} Cumulative distribution function of $n_{\rm bkg}$. 
  \textit{(right)} The total number of background sources for stars with temperatures of 5000--6000 K based on Equation (\ref{eq:bkg2}). 
  The number of observed infrared excesses are also shown as dashed lines.
  }
  \label{fig:contami}
\end{figure*}

\begin{figure}[htbp]
  \epsscale{1.15}
  \plotone{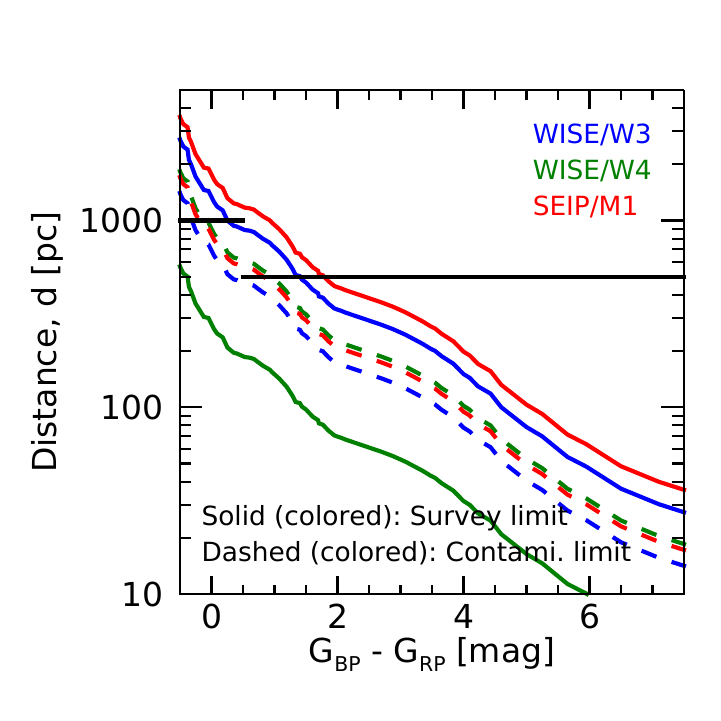}
  \caption{
  Comparison of the distance limits determined by the sensitivities of the surveys (solid lines) and those determined by the contamination (dashed lines) as a function of the color.
  The limits for $W3$, $W4$, and $M1$ are shown in blue, green, and red, respectively.
  The tentative distance thresholds adopted in Section \ref{sec:data} are shown in black lines.
  }
  \label{fig:colordistance}
\end{figure}

\begin{figure*}[htbp]
  \epsscale{1.15}
  \plotone{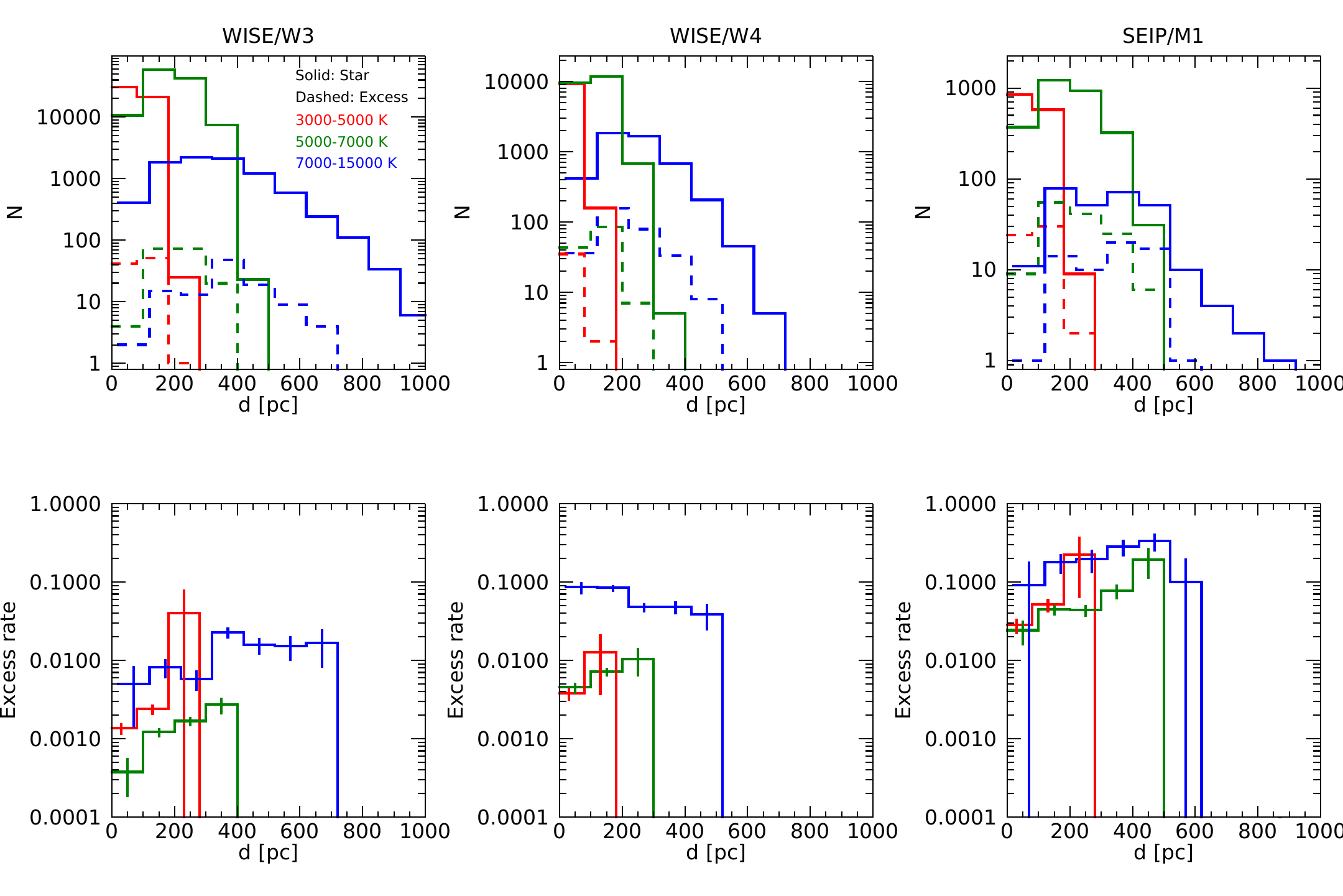}
  \caption{
\textit{(Top)} Based on Table \ref{tab3}, the distance of the target stars from the Solar System for all stars and for those with infrared excess are shown as solid and dashed lines, respectively. 
The data for $W3$, $W4$, and $M1$ are shown in blue, green, and red, respectively. 
\textit{(Bottom)} The excess rates for $W3$, $W4$, and $M1$ as a function of the distance.
  }
  \label{fig:distance}
\end{figure*}

\begin{deluxetable}{ccc}
\tablecaption{The numbers of stars and infrared excesses to derive the frequency of debris disks.}
\startdata
\tablehead{Band & $N_{\rm *}$ & $N_{\rm excess,debris}$}
W3 & 180241 & 373 \\
W4 & 36180 & 485 \\
M1 & 4587 & 255 \\
\enddata
\tablecomments{
The number of target stars and the detected infrared excesses are presented as similar to Table \ref{tab2}, but the sample is reduced by the criteria for the stellar flux and excess ratio, as discussed in Sections \ref{sec:contami} and \ref{sec:efficiency}, respectively.
}
\label{tab3}
\end{deluxetable}

\subsection{Detection efficiency of infrared excess} \label{sec:efficiency}
In the sensitivity-limited survey, the detection limit of the excess ratio varies from target to target.
Here, we introduce the excess ratio between the excess flux and the stellar flux:
\begin{equation}
  r_{\rm excess}=\frac{f_{\rm obs}-\langle c \rangle f_{\rm *,model}}{\langle c \rangle f_{\rm *,model}}.
\end{equation}
We employed the excess ratio to evaluate the detection limit of infrared excesses, because it is comparable among target stars regardless of their distance from the Solar System.
We adopted the lower limit of the detectable excess ratio, denoted by $r_{\rm excess,low}$, so that the corresponding excess was equal to $3\sigma_{\rm obs}$.
In other words, 
\begin{eqnarray}
  r_{\rm excess,low} &=& \frac{3\sigma_{\rm obs}}{\langle c \rangle f_{\rm *,model}},
\end{eqnarray}
corresponding to $\bar{\chi} = 3$.
Although Equation (\ref{eq:photsphere}) practically constrains the lower limit of the detectable excess ratio, $r_{\rm excess,low} \lesssim 1$, more stringent limits are required to obtain reliable disk frequencies with a smaller infrared excess.
The cumulative distribution functions of $r_{\rm excess,low}$, CDF($r_{\rm excess,low}$), in our sample for infrared excesses can be interpreted as the fraction of stars for which an infrared excess greater than $r_{\rm excess,low}$ can be detected.
The detection efficiency of the survey for an excess ratio $\epsilon(r_{\rm excess})$ can be expressed by the cumulative distribution function of $r_{\rm excess,low}$ as
\begin{eqnarray}
  \epsilon(r_{\rm excess}) &=& {\mathrm{CDF}(r_{\rm excess}).}
  \label{eq:efficiency}
\end{eqnarray}
Figure \ref{fig:efficiency} shows the detection efficiencies $\epsilon(r_{\rm excess})$ for $W3$, $W4$, and $M1$.
An underestimation of the disk frequency owing to the low detection efficiency for faint excess may be corrected by dividing by $\epsilon(r_{\rm excess})$.
However, this correction may be less reliable at smaller $r_{\rm excess}$ values when $\epsilon(r_{\rm excess})$ varies rapidly within a narrow range of $r_{\rm excess}$.
While $\epsilon(r_{\rm excess})$ for $W4$ gradually increases from $r_{\rm excess} = 0.1$ to 1.0, those for $W3$ and $M1$ increase steeply at $r_{\rm excess} \approx 0.2$ (see Figure \ref{fig:efficiency}).
Furthermore, because of the low $\epsilon(r_{\rm excess})$ at $r_{\rm excess} \leq 0.4$ in $W4$, the number of sources with excess in this range is so small that the disk frequencies can only be derived with large uncertainties. 
Therefore, we only discuss the frequencies of debris disks where $r_{\rm excess}$ is greater than 0.2, 0.4, and 0.2 for $W3$, $W4$, and $M1$, respectively.

\subsection{Background contamination} \label{sec:contami}
The presence of background objects (e.g., galaxies) can accidentally contaminate the flux of target stars.
The number of background objects within an aperture is expressed as follows:
\begin{eqnarray}
  n_{\rm bkg}(f_{\nu}) = \frac{dN}{df_{\nu}dA} \times \Delta f_{\nu} \times A_{\rm aper},
  \label{eq:bkg1}
\end{eqnarray}
where $dN/df_{\nu}dA$, $\Delta f_{\nu}$, and $A_{\rm aper}$ denote the differential number density per flux and solid angle $({\rm mJy}^{-1} {\rm sr}^{-1})$, the flux bin for the distribution of $n_{\rm bkg}(f_{\nu})$, and the solid angle adopted in the aperture photometry, respectively.
In this work, a circular PSF shape was assumed for this area.
The differential number density has been obtained via surveys of distant galaxies \citep[e.g.,][]{papovich04}. \par

We calculated $n_{\rm bkg}$ to investigate whether the infrared excess was caused by background sources (left in Figure \ref{fig:contami}).
For the SEIP/$M1$, we refer to Table 2 in \cite{papovich04} for an estimate of $dN/df_{\nu}dA$.
Because no data for the $dN/df_{\nu}dA$ value for WISE are available online, we employed instead the source catalog and completeness of an ${\it AKARI}$/IRC survey for the north ecliptic pole \citep{murata13}.
Because the pointed observations by {\it AKARI} are deeper than those of the WISE survey, the number density of background sources for WISE can be estimated from the {\it AKARI} data.
To reconcile the difference in the bandpass filters between {\it AKARI} and WISE, we used the averages of $dN/df_{\nu}dA$ obtained by the two broadband filters of {\it AKARI}: $S11$ and $L15$ for $W3$, $L18W$ and $L24$ for $W4$. \par

To obtain reliable frequencies of the debris disks, we must set a threshold at which the number of debris disks dominates contaminated objects.
By computing the cumulative distribution function of $n_{\rm bkg}$, we find that the contamination due to background sources fainter than 0.1--0.2 mJy can occur significantly even for field stars; the CDFs reach $\sim 1$ at 0.2 mJy for $W4$ and at 0.1 mJy for $W3$ and $M1$, respectively (middle in Figure \ref{fig:contami}). 
Here, a threshold flux of 0.1--0.2 mJy for infrared excess is defined as $f_{\rm excess,threshold}$.
In addition, a more stringent threshold should be adopted for $W3$, because the disk frequency, which can be inherently lower at short wavelengths, can be significantly overestimated owing to contamination from background sources.
Therefore, we employed a threshold at which the CDF was below $\sim 0.01$ for $W3$, or 0.6 mJy (middle in Figure \ref{fig:contami}).
In summary, we adopted 0.6, 0.2, and 0.1 mJy as $f_{\rm excess,threshold}$ for $W3$, $W4$, and $M1$, respectively. \par

By combining the criterion for excess flux with the threshold of $r_{\rm excess}$ for any statistical discussion of the frequencies of the debris disks (Section \ref{sec:efficiency}), we can determine the critical stellar flux as follows:
\begin{eqnarray}
  f_{\rm *,threshold} = \frac{f_{\rm excess,threshold}}{r_{\rm excess,threshold}},
  \label{eq:photoflux}
\end{eqnarray}
where $r_{\rm excess,threshold}$ is the lower limit of the excess ratio for a valid discussion on the disk frequencies; namely, $r_{\rm excess,threshold} =$ 0.2, 0.4, and 0.2 for $W3$, $W4$, and $M1$, respectively, as determined in Section \ref{sec:efficiency}.
Then, the threshold fluxes for stars to mitigate background contamination, $f_{\rm *,threshold}$, are 3, 0.5, and 0.5 mJy for $W3$, $W4$, and $M1$, respectively.
By adopting threshold fluxes, $\nbright$ stars in total remained as the sample for deriving the disk frequencies ([f] in Figure \ref{fig:flowchart}).
Table \ref{tab3} lists the number of stars and those with infrared excess that meet the criteria for the photospheric fluxes and excess ratios described above. 
Among the reliable samples for the frequency of debris disks, the numbers of sources with debris disks were $\ndwiset$, $\ndwisef$, and $\ndseipm$ for $W3$, $W4$, and $M1$. 
The comparison of Table \ref{tab2} and \ref{tab3} confirms the huge reduction for the number of infrared excess at $W3$ as well as its excess rate.
In the wavelength of 10--30 $\mu$m, the better sensitivity can be achieved at shorter wavelengths, resulting in infrared excess with small excess ratio can be detected.
The reduction suggests that much of the faint infrared excess at $W3$ is produced by the contamination, although it is not excluded that the faint excess simply dominates the bright excess.
In the following section, we discuss the frequencies of debris disks based on the sample shown in Table \ref{tab3}. \par

The constraints on photospheric fluxes were essentially equivalent to those on the target distance. 
Figure \ref{fig:colordistance} compares the distance limits determined by the sensitivities of the surveys and those determined by the contamination.
It is clear that the detection limit of $W4$ is determined by the survey sensitivity, while those of $W3$ and $M1$ are determined by the contamination. It can also be confirmed that the tentative distance thresholds determined in the first step (1 kpc for $G_{\rm BP} - G_{\rm RP} < 0.5 $ and 500 pc for $G_{\rm BP} - G_{\rm RP} > 0.5 $; see Section \ref{sec:data}) almost completely cover the final detection limit of the debris disks. \par

Figure \ref{fig:distance} shows the distribution of the target stars selected using Equation (\ref{eq:photoflux}) and their excess rates as a function of distance.
Although the excess rates for low-mass and intermediate-mass stars colored blue and red, respectively, showed no significant biases, the excess rates for solar-type stars with $T_{\rm eff} = 5000-7000$ K increased as a function of distance, especially for the most distant bin of $M1$ (bottom right panel of Figure \ref{fig:distance}).
For more massive stars, the search for infrared excess can be performed for more distant stars, indicating that massive stars can dominate a sample of distant stars. 
As will be shown in a later section, the disk frequencies were higher for more massive stars. 
In particular, the disk frequencies show a sudden increase at 7000 K, and the excess rate of more distant bins is expected to be higher, as shown in Figure \ref{fig:distance}.
Additionally, the number of stars in the most distant bin was not sufficiently large to affect the disk frequencies of the sample. \par

Although the contaminated sources should not dominate the sample because of the constraints on the stellar flux, some infrared excess is still due to background objects.
To further mitigate the effect of background contamination, we statistically evaluated the ratio of the number of contaminated objects to the number of others.
Assuming a spatially uniform density for $dN/df_{\nu}dA$, we can compute the number of background sources as shown in the right panel of Figure \ref{fig:contami} as follows,
\begin{eqnarray}
  N_{\rm bkg}(f_{\nu}) &=& N_{\rm obs} \times n_{\rm bkg}(f_{\nu}) \nonumber \\
  &=& N_{\rm obs} \times \frac{dN}{df_{\nu}dA} \Delta f_{\nu} \pi \left( \frac{\rm FWHM_{\rm PSF}}{2} \right)^{2},
  \label{eq:bkg2}
\end{eqnarray}
where $N_{\rm obs}$ denotes the number of stars observed.
We introduce a factor that corresponds to the observed infrared excess caused by background contamination as follows:
\begin{eqnarray}
  \begin{array}{ll}
    p_{\rm bkg}(f_{\rm excess}) = \left\{
    \begin{array}{ll}
      \frac{N_{\rm bkg}(f_{\rm excess})}{N_{\rm excess}(f_{\rm excess})} & (N_{\rm bkg} \leq N_{\rm excess}),\\
      1 & (\mathrm{otherwise})
    \end{array}
    \right.
  \end{array}
  \label{eq:contami}
\end{eqnarray}
where $f_{\rm excess}$ denotes flux of infrared excess.
Here, we simply assume that $n_{\rm bkg}$ is spatially uniform because stars in highly contaminated regions, such as the galactic plane, have already been excluded from the sample. 

\begin{figure*}[htbp]
  \epsscale{1.15}
  \plotone{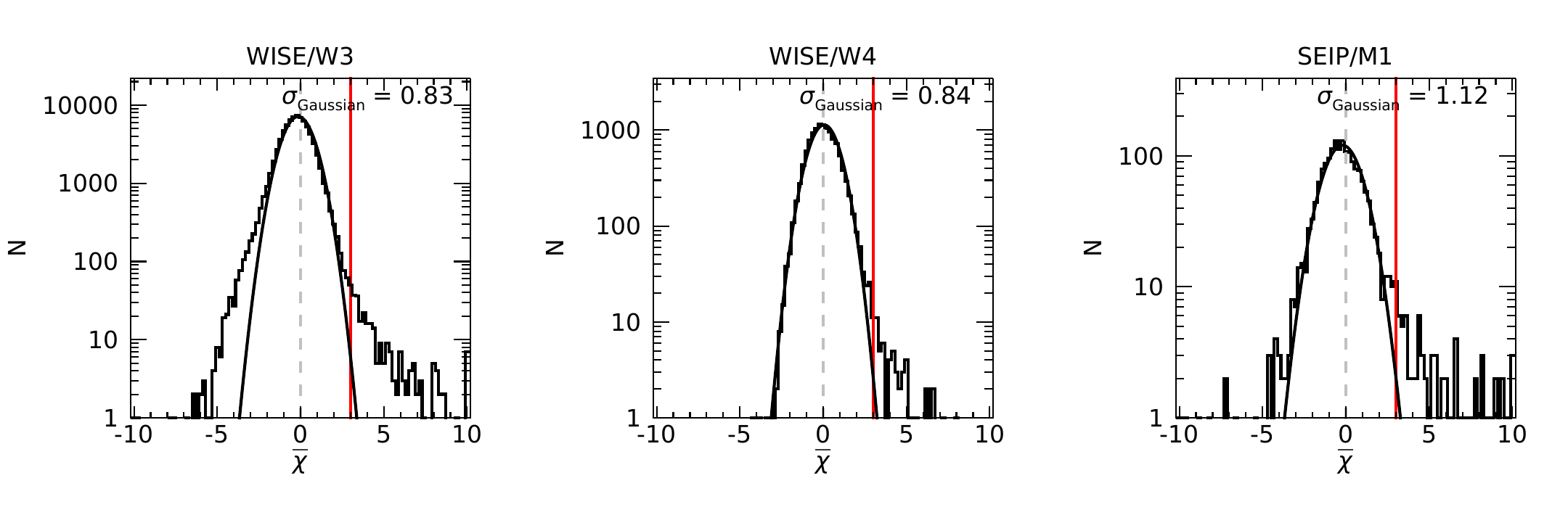}
  \caption{
  Distributions of $\bar{\chi}$ for stars with effective temperatures between 5000--6000 K. 
  In these panels, red vertical lines indicate a threshold to identify infrared excess ($\bar{\chi} > 3$).
  }
  \label{fig:astrexc}
\end{figure*}

\section{Frequency of debris disk} \label{sec:result}

\subsection{Distributions of $\bar{\chi}$} \label{sec:astrexcess}
Some of the sources should simply be due to statistical fluctuations, although we identified an infrared excess at the 3$\sigma$ confidence level. 
Indeed, Figure \ref{fig:astrexc} shows the distribution of $\bar{\chi}$ for stars with effective temperatures in the range of $5000-6000$ K, revealing large populations of negative $\bar{\chi}$. 
Here, we estimate the infrared excess due to astronomical sources by simply subtracting the number of negative excesses $N_{\rm excess}(-\bar{\chi})$ from the number of corresponding positive excesses $N_{\rm excess}(\bar{\chi})$ as follows: 
\begin{eqnarray}
  N_{\rm excess,astr}(\bar{\chi}) = N_{\rm excess}(\bar{\chi}) - N_{\rm excess}(-\bar{\chi}).
  \label{eq:efficiency_diff}
\end{eqnarray}
We introduce $p_{\rm excess,astr}$ to represent the probability that each infrared excess is caused by astronomical sources as follows:
\begin{eqnarray}
  \begin{array}{ll}
    p_{\rm excess,astr}(\bar{\chi}) = \left\{
    \begin{array}{ll}
      \frac{N_{\rm excess,astr}(\bar{\chi})}{N_{\rm excess}(\bar{\chi})} & (N_{\rm excess,astr}(\bar{\chi}) \geq 0),\\
      0 & (\mathrm{otherwise}).
    \end{array}
    \right.
  \end{array}
  \label{eq:astrexc}
\end{eqnarray}
If debris disks dominate the astronomical sources with an infrared excess, $p_{\rm excess,astr}(\bar{\chi})$ can be interpreted as the expected value for the number of debris disks by an infrared excess. \par

\subsection{Frequencies of debris disks in the sensitivity-limited survey}
To obtain the debris disk frequencies under a uniform criterion, the detection efficiency, background contamination, and statistical fluctuations must be corrected.
In the calibration-limited surveys of previous studies, the term $p_{\rm excess,astr}(\bar{\chi})$ in Equation (\ref{eq:astrexc}) has been employed to count the number of debris disks.
Here, we introduce the value $n_{\rm debris,sens}$ to represent the expected value for the number of debris disks by an infrared excess in the case of sensitivity-limited survey as, 
\begin{eqnarray} \label{eq:correction}
  n_{\rm debris,sens} &=& \left[\frac{1-p_{\rm bkg}(f_{\rm excess})}{\epsilon(r_{\rm excess})} \right] p_{\rm excess,astr}(\bar{\chi}),
\end{eqnarray}
where $\epsilon(r_{\rm excess})$ is the detection efficiency derived in Section \ref{sec:efficiency}, $p_{\rm bkg}(f_{\rm excess})$ is the probability of background contamination derived in Section \ref{sec:contami}, and $p_{\rm excess,astr}(\bar{\chi})$ is the probability that each infrared excess is caused by astronomical signals as described in Section \ref{sec:astrexcess} (see also Equations (\ref{eq:efficiency}), (\ref{eq:contami}), and (\ref{eq:astrexc})). 
Each of the three components of an object is calculated independently for some bins of $r_{\rm excess}$, $f_{\rm excess}$, or $\bar{\chi}$ and the effective temperature. \par

\begin{figure*}[htbp]
  \epsscale{1.15}
  \plotone{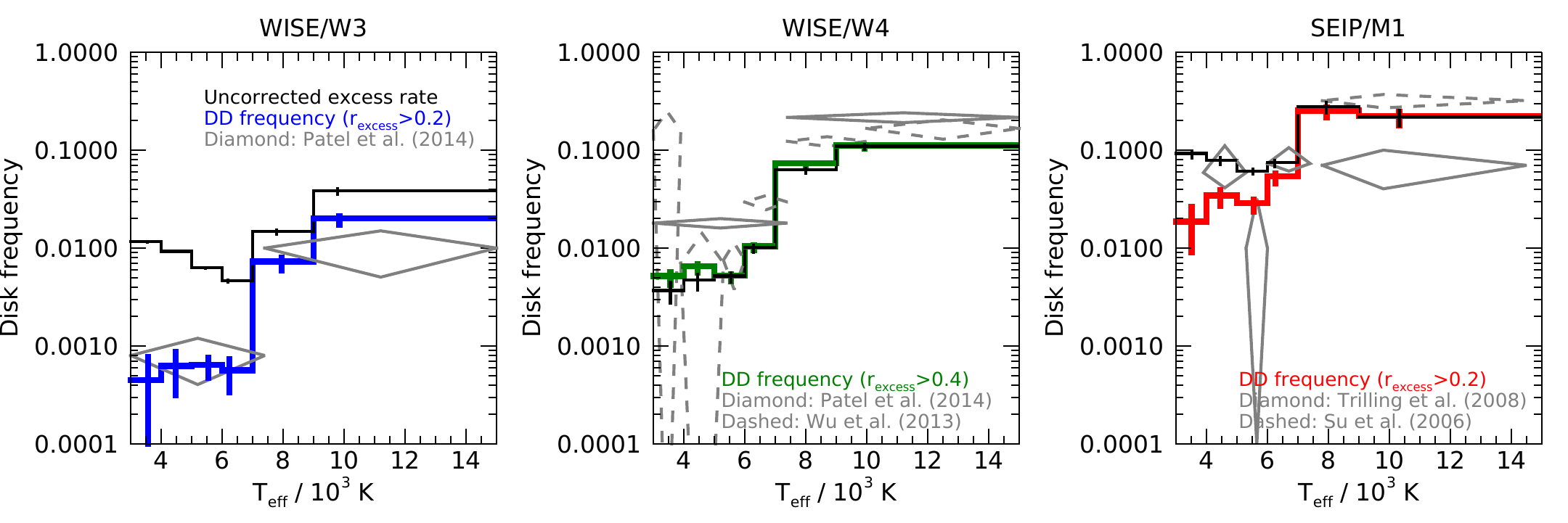}
  \caption{
  Disk frequencies are presented as a function of stellar effective temperature. 
  The colored lines show the frequency of debris disks based on all the corrections and criteria in Sections \ref{sec:correction} and \ref{sec:result}, and the black lines show the excess rate with no correction. 
  The gray diamonds indicate the effective temperature bins and the 1$\sigma$ uncertainty of the disk frequencies obtained by previous studies \citep{su06, trilling08, wu13, patel14}, for the horizontal and vertical extent, respectively.
  Due to the limited information of the sample in previous studies, particularly stars without disks, except for the intermediate-mass stars of MIPS ch1, we set the bins of effective temperature as follows; BA-type for $T_{\rm eff}$ = 7400--15000 K, F-type for 6000--7400 K, G-type for 5300--6000 K, K-type for 3900--5300 K, and M-type for 3000--3900 K.
  }
  \label{fig:excratecomp}
\end{figure*}

Figure \ref{fig:excratecomp} shows the uncorrected and corrected estimates of the disk frequencies.
The uncorrected estimate is inferred from the raw sample in Table \ref{tab2}, whereas the corrected value is based on the selected sample in Table \ref{tab3} and the correction by Equation (\ref{eq:correction}).
The uncorrected excess rate overestimates the disk frequency for $W3$ and $M1$, particularly for the cool stars.
Because the surveys by $W3$ and $M1$ were deep, they suffered from background contamination.
By contrast, the uncorrected excess rate of $W4$ slightly underestimated the disk frequency.
Although the correction for bright targets detected in $W4$ was very small, some disks may have been missed because of the low detection efficiency of the sample.
Because the number of target stars and excess stars is at least several times larger than those in previous studies (Table \ref{tab4}), this work is the most systematic survey of warm debris disks for different spectral types of host stars.

\begin{deluxetable*}{c | ccc | ccc cc}
  \tablecaption{Comparison of disk frequencies obtained by $W3$, $W4$, and $M1$}
\startdata
\tablehead{Sp. type & $N_{\rm *}$ & $N_{\rm debris}(N_{\rm excess})$ & Rate [$\%$] & $N_{\rm *,ref}$ & $N_{\rm debris,ref}$ & Rate$_{\rm ref}$ [$\%$] & Reference\tablenotemark{a}}
WISE/$W3$ \\
late-B and A & 7312 & 90.3(103) & 1.23$\pm$0.13   & & & 1.0$\pm$0.5 & P14 \\
F            & 43745 & 30.5(65) & 0.06$\pm$0.02 \\
G            & 63175 & 36.7(82) & 0.05$\pm$0.01 \\
K            & 48488 & 33.5(90) & 0.06$\pm$0.02 \\
M            & 17521 & 8.2(33) & 0.04$\pm$0.04 \\
FGK          & 155408 & 100.8(237) & 0.06$\pm$0.01 & & & 0.08$\pm$0.04 & P14 \\
\hline
WISE/$W4$ \\
late-B and A & 4200 & 384.4(287) & 9.15$\pm$0.40  & 737  & 96 & 13.0$\pm$1.3 & W13 \\
             & & &                                & & & 21.6$\pm$2.5 & P14 \\
F            & 10422 & 136.7(107) & 1.31$\pm$0.09 & 1220 & 36 &  3.0$\pm$0.5 & W13 \\
G            & 9918 & 46.3(39) & 0.46$\pm$0.06   & 531  & 4  &  0.8$\pm$0.4 & " \\
K            & 8145 & 57.0(41) & 0.70$\pm$0.07   & 142  & 1  &  0.7$\pm$0.7 & " \\
M            & 3495 & 15.8(11) & 0.45$\pm$0.09   & 19   & 3  & 15.8$_{-15.8}^{+9.1}$ & " \\
FGK          & 28485 & 240.1(187) & 0.84$\pm$0.04 & & & 1.8$\pm$0.2 & P14 \\
\hline
SEIP/$M1$ \\
late-B and A & 237 & 60.8(57) & 25.67$\pm$3.18   & 155 & & 32$\pm$5 & S06 \\
             & & &                               & 30  & &  7$\pm$3 & T08 \\
F            & 1103 & 63.8(74) & 5.78$\pm$0.83   & 110 & & 7.3$_{-1.2}^{+3.3}$ & " \\
G            & 1467 & 38.2(48) & 2.60$\pm$0.51   & 103 & & 1.0$_{-1.0}^{+2.2}$ & " \\
K            & 1232 & 42.2(56) & 3.43$\pm$0.67   & 51  & & 5.9$_{-1.8}^{+5.2}$ & " \\
M            & 548 & 10.8(20) & 1.97$\pm$0.98    & 62  & & $<$ 2.9 & " \\
FGK          & 3802 & 144.2(178) & 3.79$\pm$0.38 & 184 & & 3.8$_{-1.2}^{+1.7}$ & " \\
\enddata
\tablecomments{
The number of infrared excesses and the frequency of debris disks obtained in this study were compared with previous estimates.
The spectral types correspond to the effective temperatures as follows: BA-type for $T_{\rm eff}$ = 7400--15000 K, F-type for 6000--7400 K, G-type for 5300--6000 K, K-type for 3900--5300 K, and M-type for 3000--3900 K.
(a) The aberrations indicate the following references: S06: \cite{su06}; T08: \cite{trilling08}; W13: \cite{wu13}; and P14: \cite{patel14}.
}
\label{tab4}
\end{deluxetable*}

\begin{figure*}[htbp]
  \epsscale{1.05}
  \plottwo{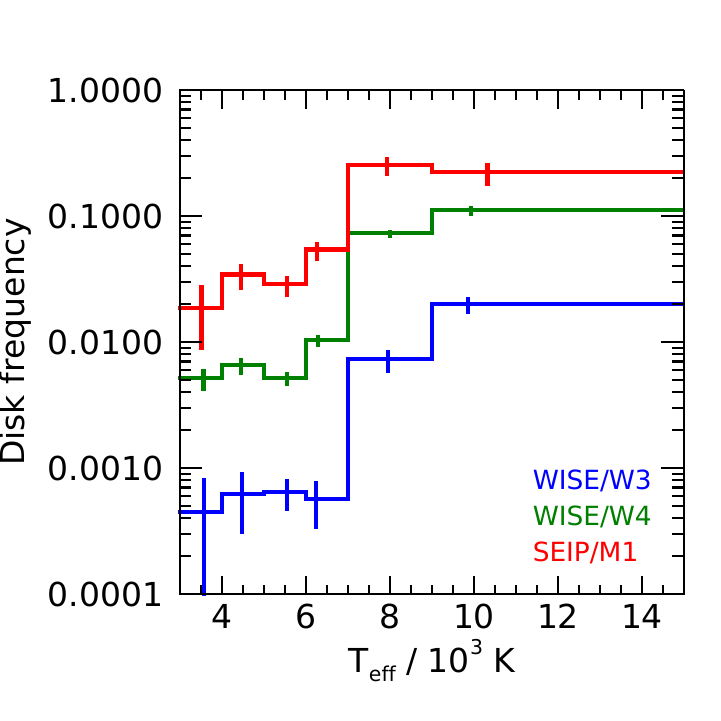}{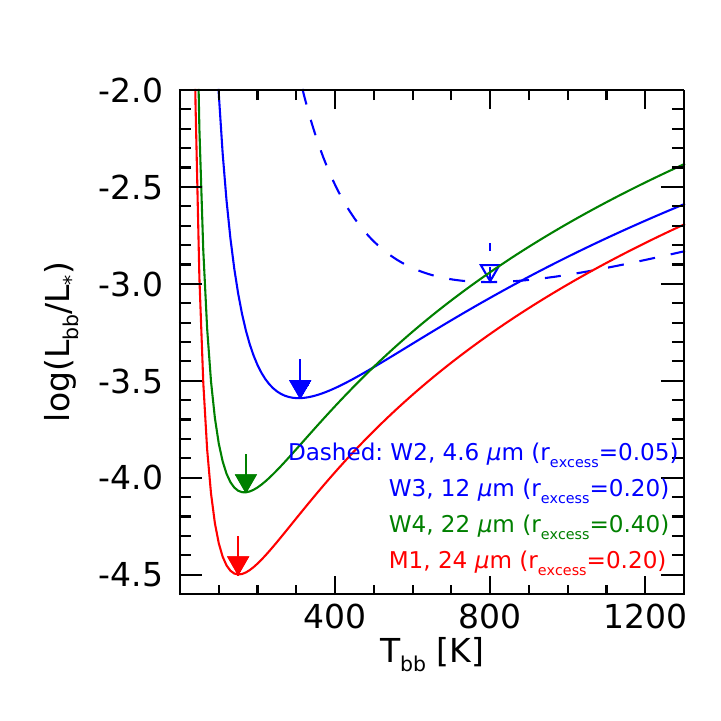}
  \caption{
  Left shows the frequencies of debris disks for the comparison among three wavelengths, based on the all the corrections and criteria in Sections \ref{sec:correction} and \ref{sec:astrexcess}.
  Right represents the detection limits of infrared excess associated with a solar-type star of $T_{\rm eff}$ = 6000 K. 
  Colors correspond to 12, 22, and 24 $\mu$m as references for WISE/$W3$, $W4$, and SEIP/$M1$, respectively. 
  The dashed line indicates the detection limit at 4.6 $\mu$m in the $W2$ band, hot disks brighter than which are not identified as infrared excesses in this work (see Section \ref{sec:diffband}). 
  The disk is considered a blackbody at a temperature of $T_{\rm bb}$.
  }
  \label{fig:ddfreq}
\end{figure*}

\subsection{Comparison with previous works}
We compare our results with those of previous studies using {\it Spitzer} or WISE data.
Comparisons of disk frequencies between different studies are complicated because they depend on several factors, such as sample selection, observational sensitivities, and the definition of excess.
Nevertheless, we find that our estimates generally agree with those of previous works, as shown in Figure \ref{fig:excratecomp}.
These minor differences can be explained by the different detection limits of the excess ratio.
\cite{wu13} identified the infrared excess of $W4$ based on infrared colors.
Their detection limits of the excess are down to 0.20--0.25 mag with 4$\sigma$ confidence, corresponding to 0.18--0.23 in $r_{\rm excess}$.
In this work, the detection limit of the infrared excess was determined by the sensitivity rather than by the systematic uncertainties of the calibration. 
Therefore, we must set a lower limit on the excess ratio to obtain the disk frequency (Section \ref{sec:efficiency}).
The lower limit of the excess ratio, $r_{\rm excess}$ = 0.4, was adopted for $W4$, resulting in a disk frequency slightly lower than the previous estimates (middle of Figure \ref{fig:excratecomp} and Table \ref{tab4}).
Based on WISE photometric colors, \cite{patel14} provided the disk frequencies of $W3$ and $W4$ with 2--4$\sigma$ confidence.
While their disk frequencies of $W4$ are also higher than those in this work, similar to the comparison with \cite{wu13}, the frequencies of $W3$ are comparable because of large uncertainties caused by lower frequencies at shorter wavelengths (left of Figure \ref{fig:excratecomp} and top in Table \ref{tab4}).
In the case of $M1$, the disk frequency was also slightly lower than or comparable to previous estimates (right of Figure \ref{fig:excratecomp} and bottom in Table \ref{tab4}).
The lower limit for $M1$ in this work was $r_{\rm excess} = 0.2$, which is similar to that in the calibration-limited survey with the {\it Spitzer}/MIPS ch1, $r_{\rm excess}\sim0.1$ \citep{su06, trilling08}.
No clear differences were found for $M1$ as a result.

\subsection{Disk frequencies in different bands} \label{sec:diffband}
All three frequencies obtained for $W3$, $W4$, and $M1$ show similar trends with respect to the stellar temperature, and the disks are more likely to be detected at longer wavelengths (left of Figure \ref{fig:ddfreq}).
We briefly examined the detectable parameter space for disks associated with solar-type stars ($T_{\rm eff} = 6000$ K) by varying the blackbody temperature and luminosity of the disks.
The right panel of Figure \ref{fig:ddfreq} shows the parameter space based on the lower limits of the excess ratio $r_{\rm excess} = 0.2, \ 0.4, \ {\rm and} \ 0.2$ set in Section \ref{sec:efficiency} and the monochromatic excess at 12, 22, and 24 $\mu$m as references for $W3$, $W4$, and $M1$, respectively.
For different spectral types of host stars, all the curves shift similarly along the fractional luminosity, because the fluxes of the main sequence stars approach the Rayleigh-Jeans limit at mid-infrared wavelengths. 
We removed the stars with excess in the near infrared (2--5 $\mu$m) from the sample to avoid excess owing to companions in binary systems in Section \ref{sec:remove}.
As a result, our survey may have missed some fractions of hot disks. 
Using the detection limit of $W2$ with 4.6 $\mu$m as the criterion for not showing binary excess, the parameter space where hot disks could not be detected in this work is shown below the dashed line on the right in Figure \ref{fig:ddfreq}. \par

If warm disks with temperatures of approximately 200--300 K or colder disks dominate the infrared excess at mid-infrared wavelengths, these disks become difficult to detect at shorter wavelengths, where stars are brighter.
Furthermore, although detecting cold disks at shorter wavelengths with $W3$ is difficult because their fluxes decrease steeply with Wien's approximation, observations with $W4$ and $M1$ can detect both cold and hot/warm disks.
Therefore, the disk frequencies of the $W3$ were significantly lower than those at longer wavelengths.
The disk frequency of the $M1$ is several times higher than that of $W4$, although their wavelengths are similar. 
This is due to the difference in the lower limits of the excess ratio $r_{\rm excess} = 0.4$ and $0.2$ for $W4$ and $M1$, respectively. 
The difference between the disk frequencies of $W4$ and $M1$ implies that a weaker infrared excess is more common; disks with $r_{\rm excess} = 0.2-0.4$ are more common than those with $r_{\rm excess} > 0.4$, which can also be seen at the individual wavelengths in Figure \ref{fig:efficiency}.

\subsection{Dual band detection of infrared excess}
\begin{figure*}[htbp]
  \epsscale{1.15}
  \plotone{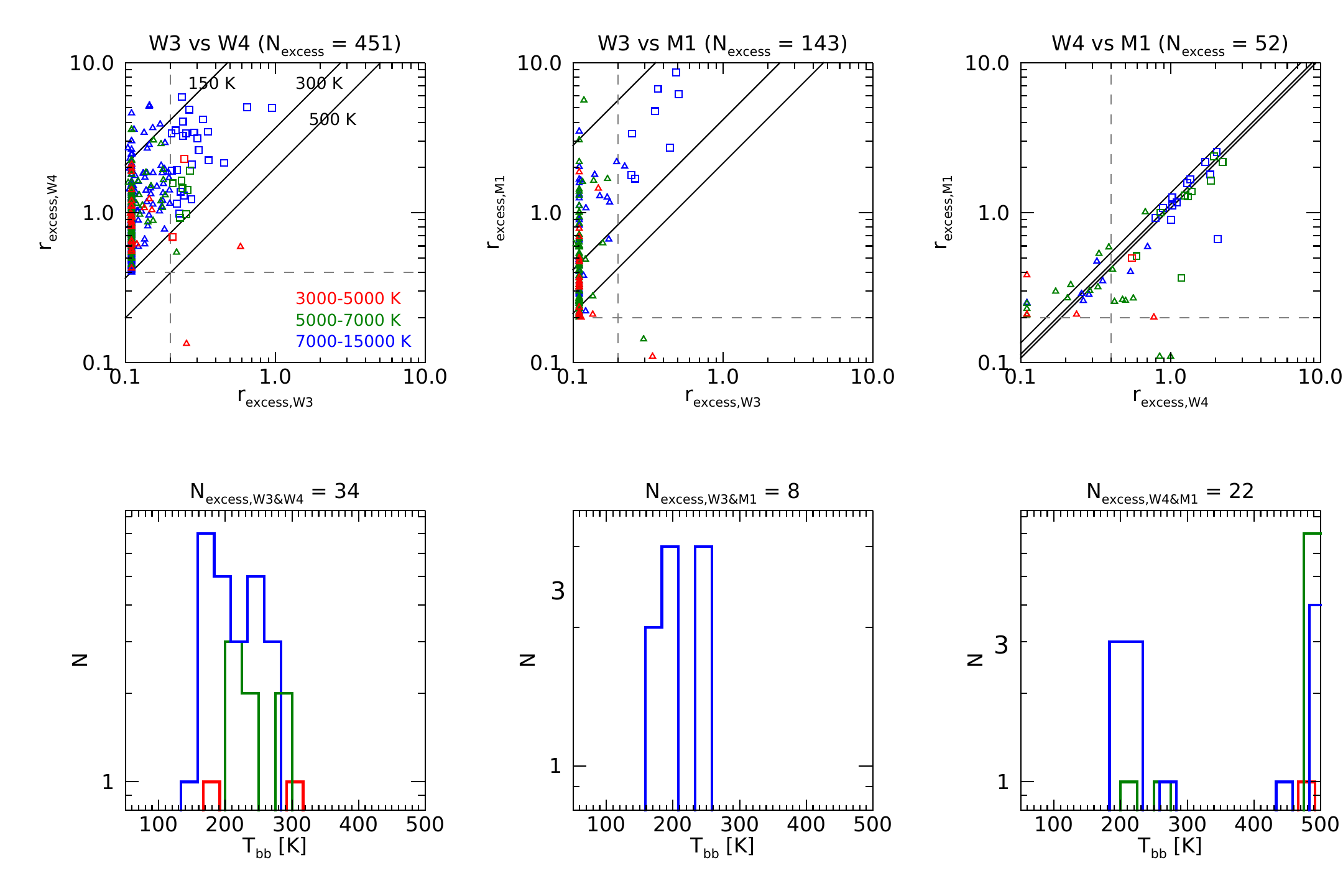}
  \caption{
  The colors of infrared excesses are presented as diagrams of excess ratios.
  The squares and triangles represent debris disks identified as infrared excesses in both two or only one wavelength band, respectively.  
  Dashed lines indicate the thresholds of excess ratio set in Section \ref{sec:efficiency}.
  The bottom half shows histograms for the blackbody temperature of debris disks whose infrared excesses are identified at two wavelength bands.
  }
  \label{fig:excesscolor}
\end{figure*}

The debris disks detected in multiple bands were also examined.
The numbers of target stars detected in two wavelength bands were $\dualexcessa$, $\dualexcessb$, and $\dualexcessc$ for the combinations of $W4/W3$, $M1/W3$, and $M1/W4$, respectively.
Table \ref{tab5} lists the number of infrared excesses detected in two wavelength bands.
Naturally, most stars with infrared excess at shorter wavelengths ($>90\%$) also exhibit excess at longer wavelengths.
However, when infrared excess was detected at longer wavelengths, less than 10 $\%$ of them showed excess at shorter wavelengths in the comparison between $W3$ and $W4$ or $M1$.
This is probably due to the lower sensitivity of $W3$ for infrared excess (right in Figure \ref{fig:ddfreq}), and it also implies that some cold disks \citep[$\lesssim$ 100 K; see, for example,][]{holland17} may be present in the sample.
Figure \ref{fig:excesscolor} shows the color of the infrared excess and the distributions of the blackbody temperatures of the disks, indicating that the typical temperature of the disks detected at both $\sim$10 and $\sim$20 $\mu$m is 150--300 K.
On comparing $W4$ and $M1$, the excess ratios were consistent (right panel in Figure \ref{fig:excesscolor}). 
However, some of the excesses in $W4$ were significantly higher than those in $M1$. 
Such a large excess of $W4$ is likely to be from background contamination owing to its poorer spatial resolution than $M1$, but intrinsic variability of the disk itself is also be a possible reason \citep[see, for example,][]{melis12, su19}.

\begin{table*}[htbp]
\caption{Dual band detection of infrared excess}

  \begin{minipage}[c]{0.32\hsize}
    \centering
    \begin{tabular}{c | cc}
$W4 \backslash W3$ & No-excess & Excess \\
\hline
Excess & 414 & 34\\
No-Excess & 34910 & 3\\
    \end{tabular}
  \end{minipage}
  \begin{minipage}[c]{0.32\hsize}
    \centering
    \begin{tabular}{c | cc}
$M1 \backslash W3$ & No-excess & Excess \\
\hline
Excess & 133 & 8\\
No-Excess & 2876 & 2\\
    \end{tabular}
  \end{minipage}
    \begin{minipage}[c]{0.32\hsize}
    \centering
    \begin{tabular}{c | cc}
$M1 \backslash W4$ & No-excess & Excess \\
\hline
Excess & 28 & 22\\
No-Excess & 1003 & 2\\
    \end{tabular}
  \end{minipage}
  
\tablecomments{
The number of infrared excesses detected in two wavelength bands. 
}
\label{tab5}
\end{table*}

\begin{figure*}[htbp]
  \epsscale{1.15}
  \centering
  \plotone{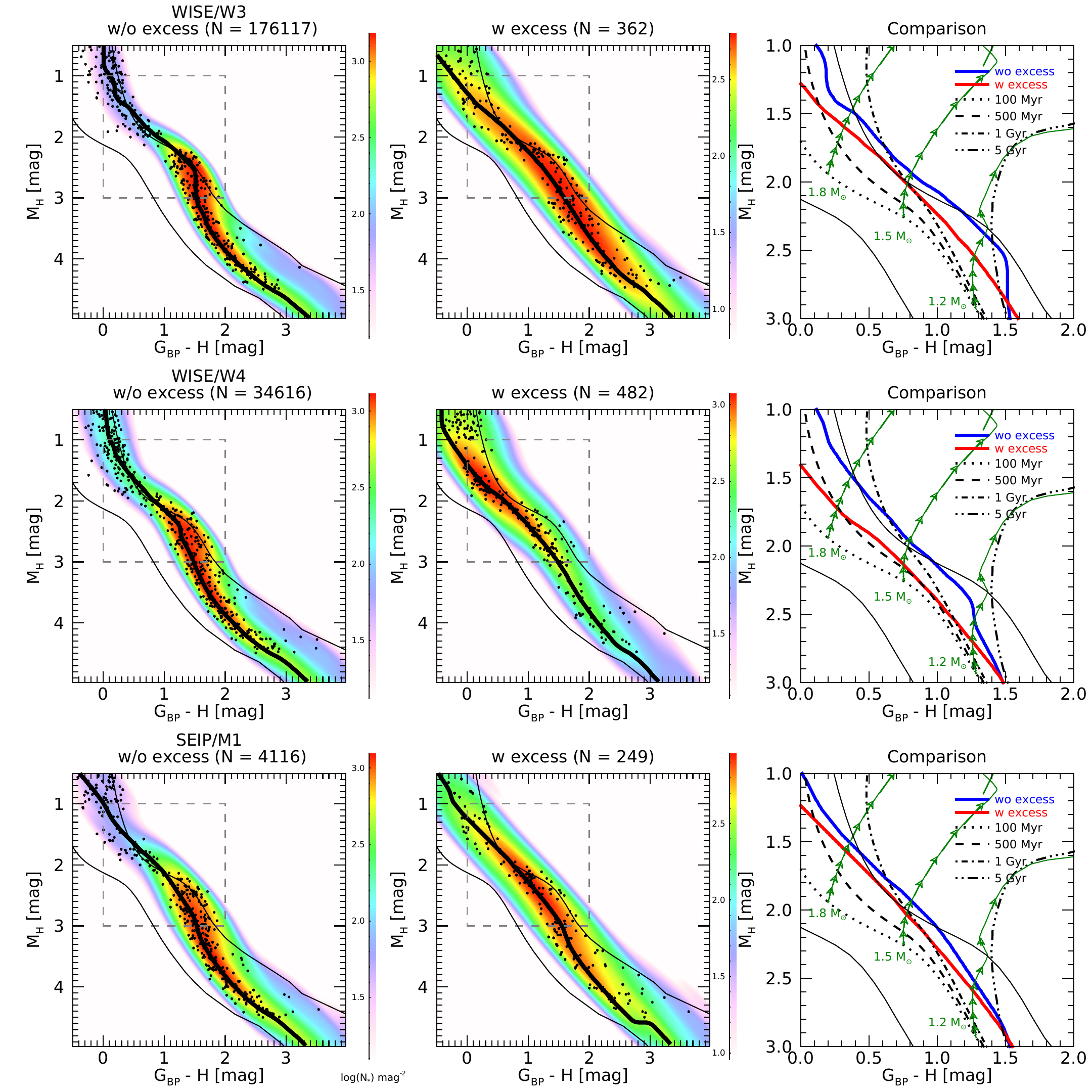}
  \caption{
  The left and middle panels represent the color-magnitude diagrams for the stars without infrared excesses and those with excesses, respectively.
  The thick black lines indicate the ridges of the density distributions.
  The thin black lines show the MS region discussed in Section \ref{sec:cmd}.
  The right panels show comparisons of the ridges with isochrones from 100 Myr to 5 Gyr and evolutionary tracks of 1.2, 1.5, and 1.8 $M_{\odot}$ stars in the zoomed area presented by dashed lines in the left and middle panels.
  }
  \label{fig:ddcmd}
\end{figure*}

\section{Discussion} \label{sec:discussion}

\subsection{Debris disks on the color-magnitude diagrams} \label{sec:ddcmd}
To estimate the ages of the sample stars, we examined their positions on color-magnitude diagrams. 
Kernel density estimation was used to compute the 2D probability density function of our sample stars from the sparse color-magnitude positions. 
We used the Python code \texttt{scipy.stats.gaussian$\_$kde}, which implements the Gaussian kernel, and the method described in \cite{scott92} for bandwidth determination.
We traced ridges of these density distributions.
The left and middle panels of Figure \ref{fig:ddcmd} present the color-magnitude diagrams for stars without and with infrared excesses, respectively.
The right panels of Figure \ref{fig:ddcmd} represent their comparison with respect to the presence of infrared excess, where isochrone curves with ages from 100 Myr to 5 Gyr and evolutionary tracks of 1.2--1.8 $M_{\odot}$ are also shown.
The stellar evolution in this age and mass range corresponds to the evolution of main sequence stars into giants.
For all three bands, particularly $W4$, we found that intermediate-mass stars with infrared excesses were bluer and fainter along each evolutionary track than those without them.
While early F- or late A-type stars, with a mass of $\sim 1.5M_{\odot}$, $G_{\rm BP}-H \sim 0.8$, and $M_{H} \sim 1.9$, are spread over some isochrones as old as approximately 1 Gyr, those with debris disks are systematically bluer or fainter, indicating that they are younger, as old as $\sim$ 500 Myr (middle row of Figure \ref{fig:ddcmd}).
The ages of low-mass stars are difficult to estimate using these diagrams owing to their very long evolutionary timescales.
Consequently, this color trend can only be observed for intermediate-mass stars. \par

The bluer color of stars with an infrared excess could have been caused by interstellar extinction and differences in stellar metallicity, both of which were ignored in the present analysis.
However, the observed color difference cannot be explained by either of these effects for the following reasons;
First, interstellar extinction makes a star appear redder and fainter, and should be more significant for a distant star. 
We did not find a clear correlation between the excess rate and distance to the Solar System, at least for low- and intermediate-mass stars (bottom panels of Figure \ref{fig:distance}). 
Secondly, metallicity can change the evolutionary track of the color-magnitude diagrams. 
Low-metal stars tend to produce bluer colors in the diagrams. 
Planet formation may tend to be suppressed in a metal-poor environment \citep[e.g.,][]{idalin04b}, whereas the planetesimals that form a debris disk may persist for a long time.
However, a color difference was observed only for intermediate-mass stars.
If the color difference is caused by differences in stellar metallicity, it should be observed not only for intermediate-mass stars, but also for low-mass stars.
Therefore, the color difference is likely a real trend that reflects the evolution of the debris disks.

\subsection{Dependence on host stars}
Our comprehensive search for the infrared excess can be used for a statistical discussion of its frequency.
As shown in Figure \ref{fig:ddfreq}, the frequency of debris disks is higher for more massive stars at higher temperatures.
We also found that the disk frequencies were almost independent of the stellar temperature for both low-mass ($T_{\rm eff}<$ 7000 K) and intermediate-mass ($T_{\rm eff}>$ 7000 K) stars, with a jump at $\sim$7000 K.
These trends can be attributed to the different ages of the target stars.
The right panel of Figure \ref{fig:cmd} shows the possible range of isochronal ages for the sample as a function of stellar mass.
The upper limits of the ages are almost independent of the stellar temperatures for both low-mass ($M_{*,{\rm initial}} < 1.0 
 \ M_{\odot}$) and intermediate-mass stars ($M_{*,{\rm initial}} > 1.5 \ M_{\odot}$), imposed by the maximum age of the Milky Way ($\sim$13.5 Gyr) and the sampling selection on the color-magnitude diagrams, respectively.
The upper limits for the stellar ages of the sample were found to change suddenly at 1.0--1.5 $M_{\odot}$, which is roughly equivalent to $T_{\rm eff} \sim 6000-7000$ K for main sequence stars.
The hotter stars should be younger because of their shorter lifetimes; therefore, disks are more likely to be detected for such young stars. \par

Debris disks may be variable and episodic objects \citep[e.g.,][]{genda15}; therefore, their evolution is difficult to understand from observations of individual objects.
While the frequency of debris disks in stars younger than 100 Myr rarely exceeds 50 $\%$ \citep[e.g.,][]{siegler07, gaspar09}, disks are also found in older stars \citep[see, for example, HD 40136 and HD 109085\footnote{$\sim$1.3 Gyrs old early F-type stars} in][]{beichman06}.
To quantify the evolution of the debris disks, we assumed that all debris disks were bright and detectable at stellar ages younger than a certain timescale determined as a function of the stellar mass.
We assume a stellar population with a uniform distribution of stellar ages from zero to their upper limits, as shown in the right panel of Figure \ref{fig:cmd}.
The timescale can be introduced as the product of the disk frequencies and the upper limits of the sample ages as follows:
\begin{eqnarray}
    t_{\rm debris} &=& f_{\rm debris} \times t_{\rm *,upper limit},
    \label{eq:ddtime}
\end{eqnarray}
where $f_{\rm debris}$ and $t_{\rm *,upper limit}$ denote the frequencies of the debris disks and upper limits of the stellar ages for the sample, respectively.
Here, we assume that the stellar age distribution is uniform but that the actual distribution is more complicated. 
A detailed study inferring the age distribution for the current sample is needed for a more rigorous discussion, but this is beyond the scope of this work. \par

The disk frequency obtained in $M1$ was used to estimate the timescale because it is the most sensitive survey for warm disks to date.
The disk frequencies are almost independent of the stellar temperatures for low-mass ($T_{\rm eff}<$ 7000 K) and intermediate-mass ($T_{\rm eff}>$ 7000 K) stars at $\sim$3 and 30 $\%$, respectively (Table \ref{tab4}).
On the other hand, the upper limits for stellar age are 13.5 and $\sim$1 Gyr for low-mass and intermediate-mass stars, respectively (right panel of Figure \ref{fig:cmd}).
The timescales of the debris disks are $\sim$400 and $\sim$300 Myr for low-mass and intermediate-mass stars, respectively. 
Although the low- and intermediate-mass stars have different disk frequencies, their timescales are nearly the same, suggesting that the difference in disk frequencies may simply reflect the difference in their stellar age distributions.
The typical age of sources with excess in the color-magnitude diagram can be estimated from the average age of stars with debris disks (Section \ref{sec:ddcmd}), where $t_{\rm debris}$ represents the duration of the debris disks. 
These two timescales agree with each other, suggesting that they are the timescales for the formation end evolution of the debris disks. \par

The evolution of debris disks has been discussed previously.
In the case of intermediate-mass stars, \cite{su06} suggested that bright warm debris disks tend to be associated with stars younger than 400 Myr.
Observations for young open clusters and associations indicate that the disk frequencies decay on timescales of 100 Myr to 1 Gyr, depending on the spectral type of the host stars \citep{siegler07, gaspar09}.
If debris disks originate from planetesimal belts such as asteroids or Kuiper belts, their brightness should decrease with time. 
For debris disks with detectable infrared excess levels, the brightness decay timescales are estimated to be $\sim 450 \ (r/1,000 \ {\rm km})^{0.96}$ Myr \citep{kobayashitanaka10, ishihara17}, where $r$ is the radius of the planetesimals. 
This estimate, with $r \ \sim$ 1000 km, supports the disk frequency evolution obtained in this work. \par

The typical radius of planetesimals is uncertain. 
In the Solar System, this radius is inferred to be $\sim$ 100 km from the size distribution of the main belt asteroids and Kuiper belt objects \citep[][and references therein.]{kobayashi16}. 
The decay timescale for the case of a planetesimal of a typical size of $\sim$ 100 km is much shorter than the ages of stars in debris disk sample, and hence detectable debris disks are not maintained in such planetesimal belts. 
Chaotic dynamic events on a timescale of hundreds of Myr are required to explain disk frequencies. 
In the Solar System, an era of significant collisional events in the first several hundred Myr has been proposed as late heavy bombardment \citep[LHB: e.g.,][]{gomes05, strom05}.
The estimated timescale for the debris disks around low-mass stars, $\sim$ 400 Myr, is comparable to the period of the LHB.
If many of the bright, warm debris disks are indeed young ($<$ 1 Gyr), our results suggest that collisional events like the LHB are relatively common among young main sequence stars. \par

To understand what determines the frequency of debris disks, stellar mass and/or age, characterization of individual target stars is essential as future work.
An empirical approach using the spin period of stars, namely gyrochronology \citep[e.g.,][]{barnes07, mamajekhillenbrand08, angus15, angus19}, may be suitable for estimating the ages of many stars.
Because Transiting Exoplanet Survey Satellite observes almost the entire sky \citep[TESS,][]{tess}, it provides the light curves of nearby stars.
Although this work simply investigates the locations of stars in the color-magnitude diagram to discuss the ages, a more rigorous analysis for inferring the age and mass distribution of the sample will be helpful for constraining the disk timescales \citep[e.g.,][]{masuda22}.
We constructed a large sample of debris disks, many of which were newly identified, particularly those around low-mass stars.
The sample can help characterize the mineral composition of disks by spectroscopy with James Webb Space Telescope/Mid-Infrared Instrument \citep[JWST/MIRI,][]{miri} and reveal the architecture of planetary systems with exoplanet surveys.

\section{Summary} \label{sec:summary}
Based on stars in the {\it Gaia} catalog and photometric data from infrared archives, we found thousands of stars with infrared excesses.
The observed excess rate derived from the entire sample was biased by several factors, such as detection efficiency of infrared excess, background contamination, and statistical fluctuations.
By correcting these biases and employing bright stars that were relatively free from contamination, we obtained the frequency of debris disks as a function of stellar effective temperature.
The frequency is roughly in agreement with previous estimates, with a few minor inconsistencies owing to the different detection limits for the infrared excess.
We confirm that more massive stars tend to show higher frequencies of debris disks.
Furthermore, the disk frequency curve as a function of stellar effective temperature is relatively flat for both low- and intermediate-mass stars, with a jump at 7000 K for all three wavelength bands: $W3$, $W4$, and $M1$.
Using color-magnitude diagrams, we find that intermediate-mass stars with infrared excess are bluer and fainter along the stellar evolutionary tracks than those without.
This can be interpreted as a difference in stellar age between the two groups; the age of intermediate-mass stars with debris disks appears to be as young as 500 Myr, whereas those without disks are at least as old as 1 Gyr.
We estimate the duration timescale of the debris disks to be 300--400 Myr, suggesting that the difference in disk frequencies between low- and intermediate-mass stars may reflect the difference in the upper limits of stellar ages rather than the stellar mass.
To understand what determines the frequency of debris disks, stellar mass and/or age, characterization of individual target stars, especially age dating, is essential and should be pursued in future work.


\software{IDL, Python 3 \citep{python3}, NumPy \citep{numpy}, SciPy \citep{scipy}, and Matplotlib \citep{matplotlib}}

\section*{Acknowledgments}
The authors thank the anonymous referee for their useful comments, and Kento
Masuda for fruitful discussions on the characterization of stellar age.
This work was supported by JSPS KAKENHI grant numbers JP22H00179, JP17H01103, JP22H01274, JP22H01278, JP18H05441, and JP21K03642.
M.A. is supported by Special Postdoctoral Researcher Program at RIKEN.
Additional data analyses were done using IDL (Exelis Visual Information Solutions, Boulder, Colorado).
We would like to thank Editage (\url{www.editage.jp}) for English language editing. \par
This work has made use of data from the European Space Agency (ESA) mission {\it Gaia} (\url{https://www.cosmos.esa.int/gaia}), processed by the {\it Gaia} Data Processing and Analysis Consortium (DPAC, \url{https://www.cosmos.esa.int/web/gaia/dpac/consortium}). Funding for the DPAC has been provided by national institutions, in particular the institutions participating in the {\it Gaia} Multilateral Agreement. 
This publication makes use of data products from the Wide-field Infrared Survey Explorer, which is a joint project of the University of California, Los Angeles, and the Jet Propulsion Laboratory/California Institute of Technology, funded by the National Aeronautics and Space Administration. 
This work is based on archival data obtained with the Spitzer Space Telescope, which was operated by the Jet Propulsion Laboratory, California Institute of Technology under a contract with NASA. Support for this work was provided by an award issued by JPL/Caltech. 
The Pan-STARRS1 Surveys (PS1) and the PS1 public science archive have been made possible through contributions by the Institute for Astronomy, the University of Hawaii, the Pan-STARRS Project Office, the Max-Planck Society and its participating institutes, the Max Planck Institute for Astronomy, Heidelberg and the Max Planck Institute for Extraterrestrial Physics, Garching, The Johns Hopkins University, Durham University, the University of Edinburgh, the Queen's University Belfast, the Harvard-Smithsonian Center for Astrophysics, the Las Cumbres Observatory Global Telescope Network Incorporated, the National Central University of Taiwan, the Space Telescope Science Institute, the National Aeronautics and Space Administration under Grant No. NNX08AR22G issued through the Planetary Science Division of the NASA Science Mission Directorate, the National Science Foundation Grant No. AST-1238877, the University of Maryland, Eotvos Lorand University (ELTE), the Los Alamos National Laboratory, and the Gordon and Betty Moore Foundation. 
The national facility capability for SkyMapper has been funded through ARC LIEF grant LE130100104 from the Australian Research Council, awarded to the University of Sydney, the Australian National University, Swinburne University of Technology, the University of Queensland, the University of Western Australia, the University of Melbourne, Curtin University of Technology, Monash University and the Australian Astronomical Observatory. SkyMapper is owned and operated by The Australian National University's Research School of Astronomy and Astrophysics. The survey data were processed and provided by the SkyMapper Team at ANU. The SkyMapper node of the All-Sky Virtual Observatory (ASVO) is hosted at the National Computational Infrastructure (NCI). Development and support of the SkyMapper node of the ASVO has been funded in part by Astronomy Australia Limited (AAL) and the Australian Government through the Commonwealth's Education Investment Fund (EIF) and National Collaborative Research Infrastructure Strategy (NCRIS), particularly the National eResearch Collaboration Tools and Resources (NeCTAR) and the Australian National Data Service Projects (ANDS). 
This research was made possible through the use of the AAVSO Photometric All-Sky Survey (APASS), funded by the Robert Martin Ayers Sciences Fund and NSF AST-1412587. 
This publication makes use of data products from the Two Micron All Sky Survey, which is a joint project of the University of Massachusetts and the Infrared Processing and Analysis Center/California Institute of Technology, funded by the National Aeronautics and Space Administration and the National Science Foundation. 
Funding for Rave has been provided by: the Leibniz Institute for Astrophysics Potsdam (AIP); the Australian Astronomical Observatory; the Australian National University; the Australian Research Council; the French National Research Agency; the German Research Foundation (SPP 1177 and SFB 881); the European Research Council (ERC-StG 240271 Galactica); the Istituto Nazionale di Astrofisica at Padova; The Johns Hopkins University; the National Science Foundation of the USA (AST-0908326); the W. M. Keck foundation; the Macquarie University; the Netherlands Research School for Astronomy; the Natural Sciences and Engineering Research Council of Canada; the Slovenian Research Agency; the Swiss National Science Foundation; the Science $\&$ Technology FacilitiesCouncil of the UK; Opticon; Strasbourg Observatory; and the Universities of Basel, Groningen, Heidelberg and Sydney. This work has made use of data from the European Space Agency (ESA) mission Gaia(\url{https://www.cosmos.esa.int/gaia}), processed by the GaiaData Processing and Analysis Consortium (DPAC, \url{https://www.cosmos.esa.int/web/gaia/dpac/consortium}). Funding for the DPAC has been provided by national institutions, in particular the institutions participating in the Gaia Multilateral Agreement. 
Guoshoujing Telescope (the Large Sky Area Multi-Object Fiber Spectroscopic Telescope LAMOST) is a National Major Scientific Project built by the Chinese Academy of Sciences. Funding for the project has been provided by the National Development and Reform Commission. LAMOST is operated and managed by the National Astronomical Observatories, Chinese Academy of Sciences. 
This research has made use of the SIMBAD database, operated at CDS, Strasbourg, France. 

\appendix
\section{Possibilities to show infrared excess by a companion in binary systems} \label{sec:app}

\begin{figure}[htbp]
  \epsscale{1.1}
  \plotone{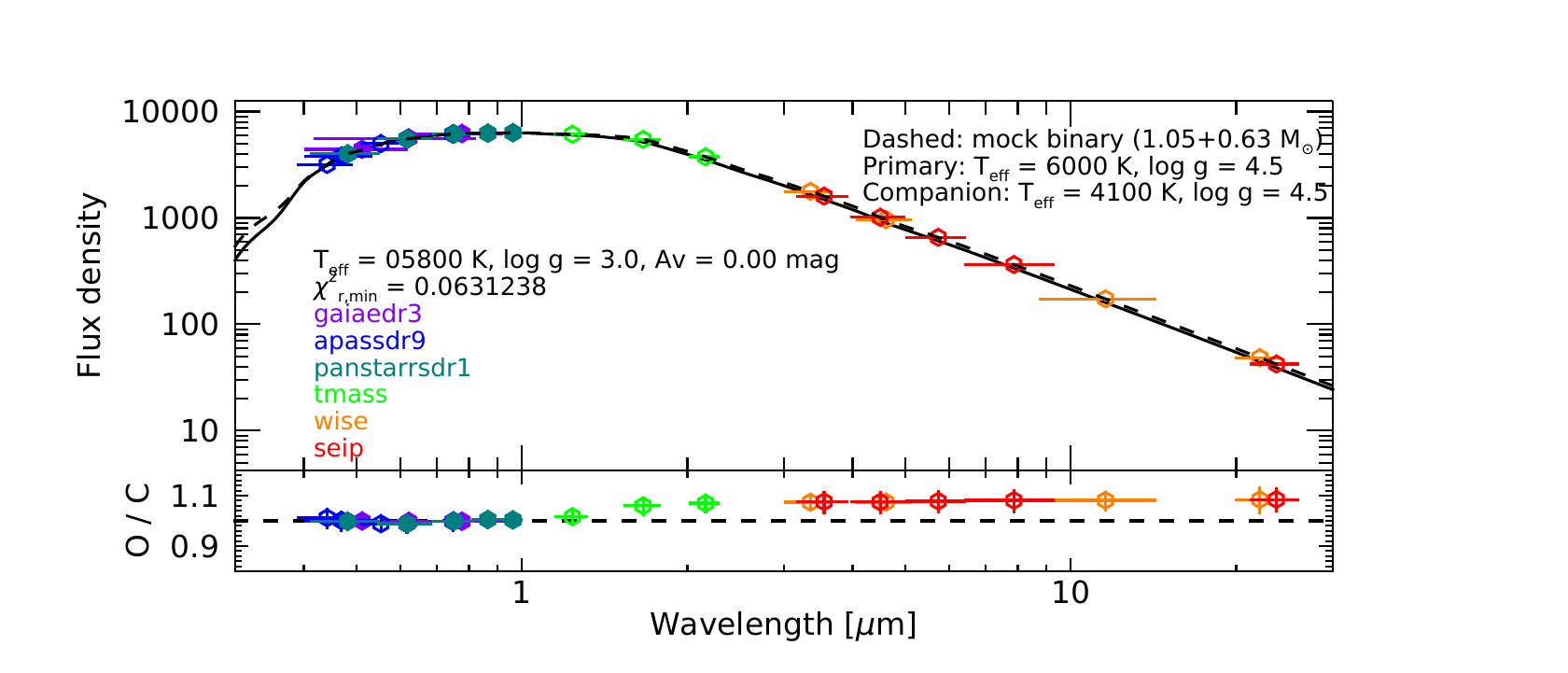}
  \caption{
  The SED fitting for a mock binary is presented, in which only photometric data with filled symbols are fitted to determine stellar parameters.
  }
  \label{fig:app_mocksed}
\end{figure}

\begin{figure}[htbp]
  \epsscale{1.1}
  \plotone{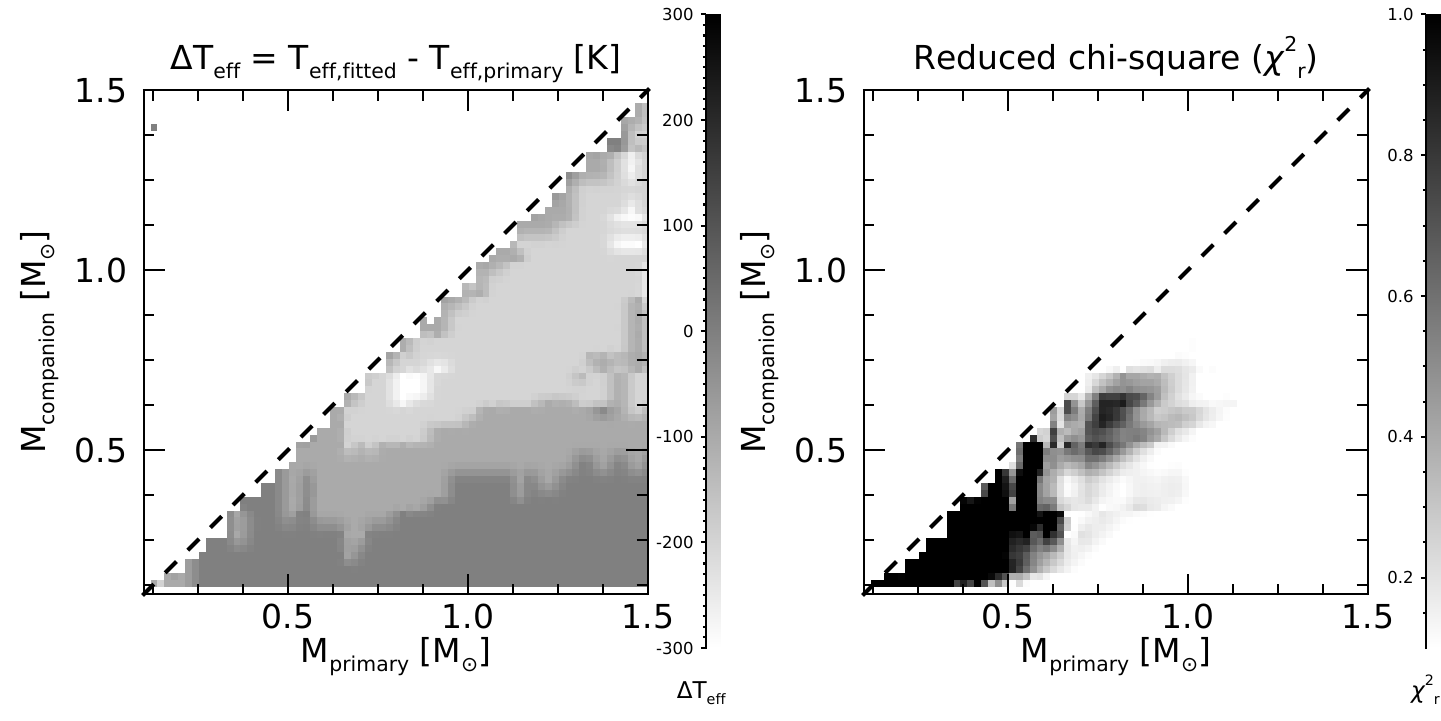}
  \caption{
  The left panel shows differences in effective temperature between the best-fit models and inputs of mock binaries. 
  The right panel presents the reduced chi-square map of the SED fitting for mock binaries.
  }
  \label{fig:app_chisq}
\end{figure}

\begin{figure}[htbp]
  \epsscale{1.1}
  \plotone{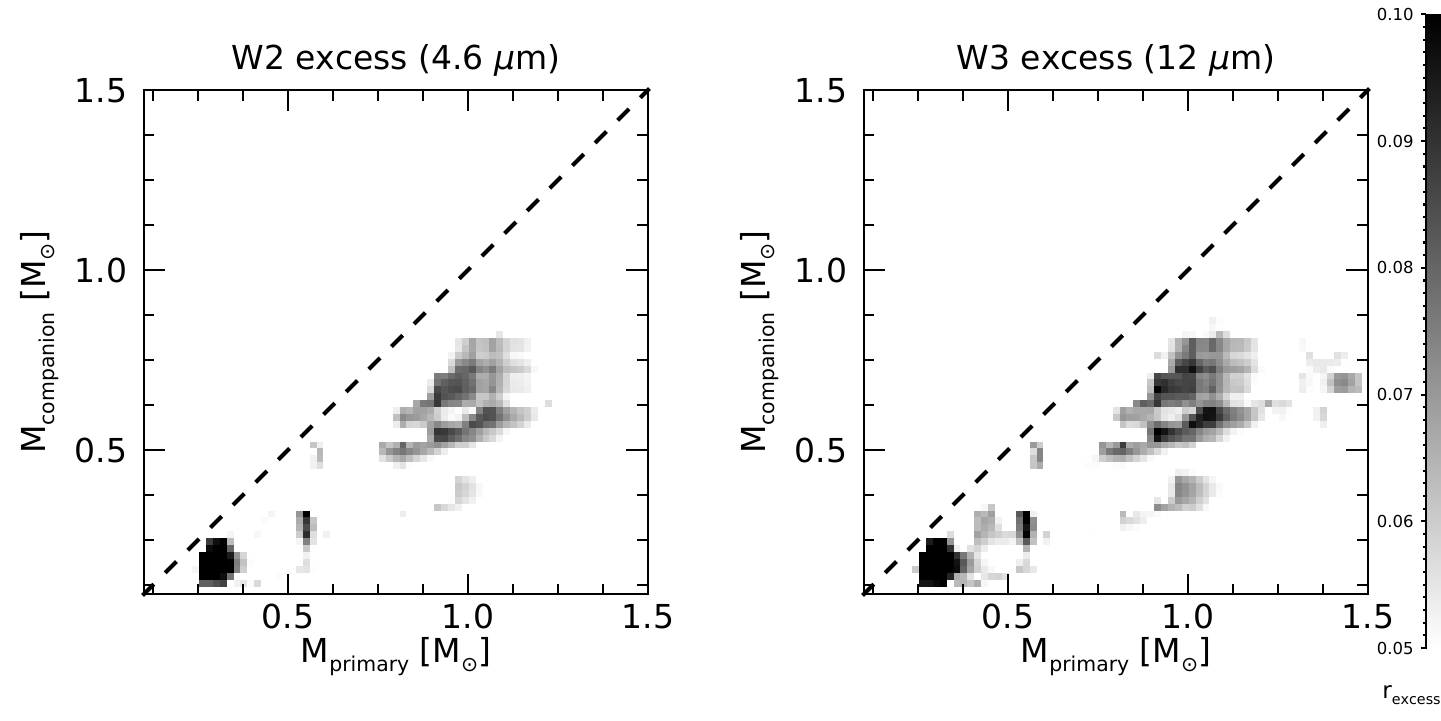}
  \caption{The maps for the excess ratios of binary excesses are presented.}
  \label{fig:app_excess}
\end{figure}

In a binary system, the mid-infrared flux ratio between the primary and its companion is nearly constant at longer wavelengths owing to the Rayleigh-Jeans approximation. 
If the stellar SED is fitted only at optical wavelengths, the infrared flux of the companion may be misinterpreted as an excess. 
To demonstrate the possibility of showing infrared excess due to companions in binary systems, we generated the SEDs of binaries based on the BT-Settle and BT-Nextgen stellar atmospheric models as mock-observational SEDs. 
Mock photometric fluxes were obtained by convolving the mock SEDs with filter response functions.
The setups of the photometric systems for {\it Gaia} eDR3/$G,G_{\rm BP},G_{\rm RP}$, APASS DR9/$g',r',i'$, and PanSTARRS DR1/$g_{\rm P1}, r_{\rm P1}, i_{\rm P1}, z_{\rm P1}, y_{\rm P1}$ were used for the demonstration.
Calibration errors in the photometric systems and possible systematic uncertainties (see Section \ref{sec:flags}) were adopted as uncertainties in the mock fluxes.
The mock fluxes were fitted using the same atmospheric models for an isolated star as described in Section \ref{sec:sed}. 
As an example, Figure \ref{fig:app_mocksed} shows the SED fitting for a mock binary composed of a 1.05 $M_{\odot}$ primary and a 0.63 $M_{\odot}$ companion with the effective temperatures of 6000 and 4100 K, respectively.
While the SED of the mock binary is similar to that of an isolated star with 5800 K at optical wavelengths, approximately 10 $\%$ infrared excess is shown at wavelengths longer than 2 $\mu$m.
For various combinations of a primary and its companion, Figure \ref{fig:app_chisq} shows the difference in the effective temperature between the primary of mock binaries and the best-fitted model of isolated stars (left) and the reduced chi-square map as a result of the fitting (right).
Similarly, Figure \ref{fig:app_excess} shows the excess ratio of the binary excess due to the failure of the extrapolation for $W2$ and $W3$. \par

Combinations of primaries and companions are important both for miscategorization as isolated stars and for showing binary excess.
In the case of binaries composed of G-type primaries and KM-type companions ($\sim$1.0 and 0.6$M_{\odot}$), although their SEDs at shorter wavelengths are well fitted as isolated stars (Figure \ref{fig:app_chisq}), they show infrared excess without significant wavelength dependence, with up to approximately $r_{\rm excess}$ = 0.1 at the $W2$ and $W3$ bands (Figure \ref{fig:app_excess}).
Intermediate-mass primaries ignore the faint fluxes from low-mass companions.
On the other hand, KM-dwarfs have complicated structures in their SEDs owing to molecular absorption; therefore, the SEDs of low-mass binaries cannot be misinterpreted as isolated stars.
The failure of extrapolation at longer wavelengths also depends on the number and wavelength of the fitted photometric data.
For example, the binary excess of G-KM binaries is not confirmed by adding the mock data of 2MASS for fitting because near-infrared fluxes are more weighted to determine the stellar SED.
However, in such cases, the reduced chi-square increases and discrepancies in the models for isolated stars cannot be ignored. \par

\begin{figure*}[htbp]
  \epsscale{1.1}
  \plottwo{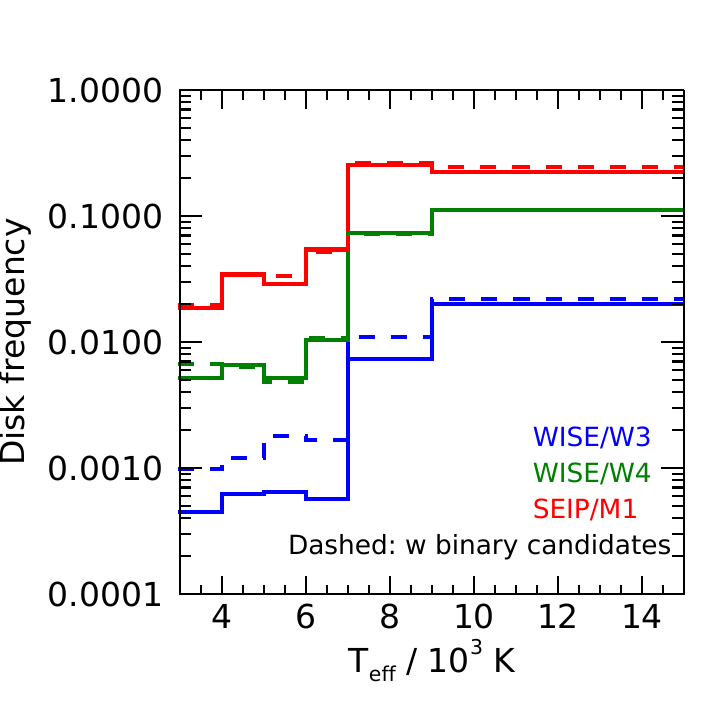}{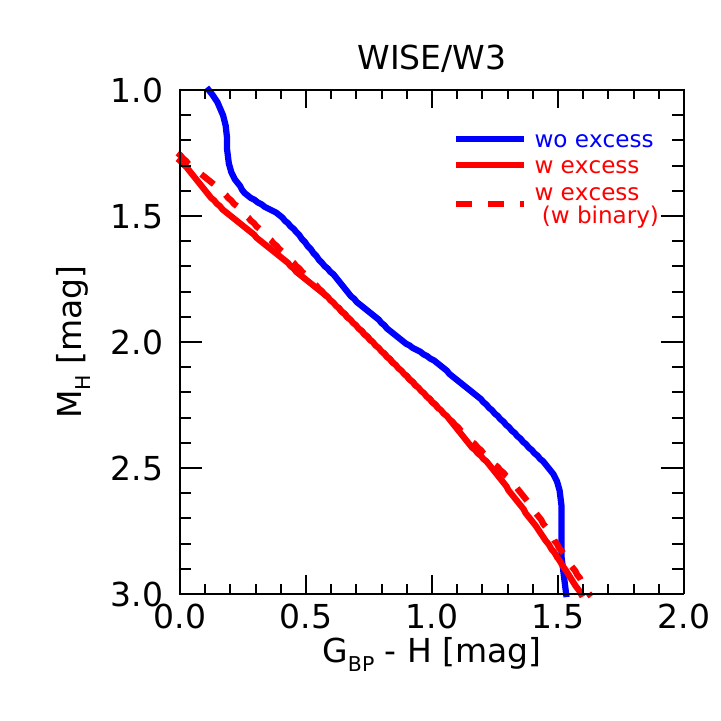}
  \caption{
  The left panel presents the frequencies of debris disks. 
  The right panel shows the ridges on the color-magnitude diagram with respect to the presence of the infrared excesses of $W3$. 
  The dashed lines represent the frequencies and the ridges, for which binary candidates showing near-infrared excesses are included in the sample.
  }
  \label{fig:app_wbincan}
\end{figure*}

Here, we compare the results with respect to the binary candidates included in the sample (see also Section \ref{sec:remove}).
The left panel of Figure \ref{fig:app_wbincan} shows the frequencies of the debris disk, similar to those in Figure \ref{fig:ddfreq}.
Although no clear differences were found in the frequencies of the debris disks in the $W4$ and $M1$ bands, the frequency of the $W3$ for solar-type stars was overestimated by the addition of binary candidates.
Because mid-infrared observations have better sensitivity at shorter wavelengths, the small excess due to a companion is more likely to be detected in $W3$ than in $W4$ or $M1$.
Similar to the right panel of Figure \ref{fig:ddcmd}, the right panel of Figure \ref{fig:app_wbincan} presents a color-magnitude diagram focused on intermediate-mass stars, where little difference is found due to binary candidates.

\bibliography{bibliography}{}
\bibliographystyle{aasjournal}

\end{document}